\documentclass[aip]{revtex4}

\usepackage{amsmath}
\usepackage{amssymb}
\usepackage{stmaryrd}

\usepackage{graphicx}

\usepackage{epsfig}

\def\ov#1{\overline{#1}}

\def\wt#1{\widetilde{#1}}
\def\vb#1{\mbox{\boldmath$#1$}}
\def\pd#1#2{\frac{\partial #1}{\partial #2}}
\def\fd#1#2{\frac{\delta #1}{\delta #2}}
\def\wh#1{\widehat{#1}}
\def\bdot{\,\vb{\cdot}\,}
\def\btimes{\,\vb{\times}\,}

\def\bhat{\wh{{\sf b}}}
\def\cal#1{\mathcal{#1}}

\def\bhat{\wh{{\sf b}}}
\def\exd{{\sf d}}

\newcommand{\bc}{\begin{center}}
\newcommand{\ec}{\end{center}}
\newcommand{\bt}{\begin{tabbing}}
\newcommand{\et}{\end{tabbing}}
\newcommand{\be}{\begin{eqnarray*}}
\newcommand{\ee}{\end{eqnarray*}}
\newcommand{\bs}{\begin{slide}}
\newcommand{\es}{\end{slide}}

\begin{document}

\title{Variational principle for the parallel-symplectic representation \\ of electromagnetic gyrokinetic theory}

\author{Alain J.~Brizard}
\affiliation{Department of Physics, Saint Michael's College, Colchester, VT 05439, USA}

\date{March 27, 2017}

\begin{abstract}
The nonlinear (full-$f$) electromagnetic gyrokinetic Vlasov-Maxwell equations are derived in the parallel-symplectic representation from an Eulerian gyrokinetic variational principle. The gyrokinetic Vlasov-Maxwell equations are shown to possess an exact energy conservation law, which is derived by Noether method from the gyrokinetic variational principle. Here, the gyrocenter Poisson bracket and the gyrocenter Jacobian contain contributions from the perturbed magnetic field. In the full-$f$ formulation of the gyrokinetic Vlasov-Maxwell theory presented here, the gyrocenter parallel-Amp\`{e}re equation contains a second-order contribution to the gyrocenter current density that is derived from the second-order gyrocenter ponderomotive Hamiltonian.
\end{abstract}


\maketitle

\section{Introduction}

The Hamiltonian formulation of the gyrokinetic Vlasov equation is based on the existence of a gyrocenter Hamiltonian function $H_{\rm gy}$ and a gyrocenter Poisson bracket $\{\;,\;\}_{\rm gy}$ in terms of which the characteristics of the gyrokinetic Vlasov equation are expressed as $\{ \ov{\cal Z}^{a},\; H_{\rm gy}\}_{\rm gy}$. When electromagnetic effects are included into the gyrokinetic formalism \cite{HLB_1988,Brizard_1989,Brizard_PhD}, the perturbed magnetic field can either appear in the gyrocenter Hamiltonian alone (in the Hamiltonian representation), in the gyrocenter Poisson bracket (in the symplectic representation), or both. These representations are said to be equivalent in the sense that the gyrocenter magnetic moment $\ov{\mu} \equiv \ov{J}\Omega/B$ is the same in all representations \cite{Brizard_Hahm_2007}.  Here, the gyrocenter phase-space coordinates $\ov{Z}^{\alpha} = (\ov{\bf X}, \ov{p}_{\|}, \ov{J}, \ov{\zeta})$ include the gyrocenter position $\ov{\bf X}$, the gyrocenter parallel (kinetic) momentum $\ov{p}_{\|}$, the gyrocenter gyroaction $\ov{J}$ and its canonically-conjugate gyroangle $\ov{\zeta}$. In the present work, because of the time-dependence of the electromagnetic-field fluctuations, we also use the extended gyrocenter phase-space coordinates $\ov{\cal Z}^{a} = (\ov{Z}^{\alpha}, \ov{w},t)$, which include time $t$ and the canonically-conjugate energy $\ov{w}$. 

\subsection{Parallel-symplectic representation}

In the parallel-symplectic representation of gyrokinetic theory \cite{Brizard_Hahm_2007}, the gyrocenter Poisson bracket $\{\;,\;\}_{\rm gy}$ contains terms involving the perturbed parallel magnetic vector potential $A_{1\|} \equiv \bhat_{0}\bdot{\bf A}_{1}$, where the nonuniform background magnetic field ${\bf B}_{0} \equiv B_{0}\,\bhat_{0}$ is expected to satisfy the standard guiding-center ordering \cite{Cary_Brizard_2009}. The perpendicular components ${\bf A}_{1\bot} (\equiv {\bf A}_{1} - A_{1\|}\,\bhat_{0})$ of the perturbed magnetic vector potential, which are associated with the parallel (compressional) component of the perturbed magnetic field $(B_{1\|} \equiv \bhat_{0}\bdot{\bf B}_{1}$), appear in the gyrocenter Hamiltonian $H_{\rm gy}$. We note that, while the so-called cancellation problem is avoided in the parallel-symplectic representation \cite{Chen_Parker_2001,Mishchenko_2004}, the parallel component of the perturbed inductive electric field appears explicitly in the gyrocenter parallel force equation, which is now expressed in terms of the gyrocenter kinetic momentum $\ov{p}_{\|}$.

The self-consistent inclusion of gyrokinetic effects into the Maxwell equations relies on the existence of a variational principle. Lagrangian \cite{Sugama_2000} and Eulerian \cite{Brizard_2000a,Brizard_2000b} variational principles for the electromagnetic gyrokinetic Vlasov-Maxwell equations were previously given in the Hamiltonian representation. These variational principles were given in the full-$f$ version of nonlinear gyrokinetic theory, in which the gyrocenter Vlasov distribution is not decomposed into a reference (time-independent) distribution function and a time-dependent perturbed distribution function. The Eulerian variational principle for the truncated-$f$ (i.e., $\delta f$) version of the nonlinear gyrokinetic Vlasov-Maxwell equations, which is the version most commonly implemented in numerical gyrokinetic simulations, were presented in Refs.~\cite{Brizard_2010,Tronko_2016}. 

The purpose of the present work is to present the Eulerian variational principle for the (full-$f$) electromagnetic gyrokinetic Vlasov-Maxwell equations in the parallel-symplectic representation, in which the parallel component of the perturbed magnetic vector potential appears in the gyrocenter Poisson bracket and the gyrocenter Jacobian. 

\subsection{Organization}

The remainder of this paper is organized as follows. In Sec.~\ref{sec:2}, we define the gyrocenter phase-space transformation used to construct the gyrocenter Poisson bracket in the parallel-symplectic representation.  Here, the gyrocenter phase-space transformation from the guiding-center phase-space coordinates ${\cal Z}^{a}$ to the gyrocenter phase-space coordinates $\ov{\cal Z}^{a}$ is derived up to second order in the perturbation ordering parameter $\epsilon$, where the time coordinate $t$ is unaffected by the transformation. We easily verify that our transformation yields a gyrocenter gyroaction $\ov{J}$ that is independent of the representation used in gyrokinetic theory, i.e., the gyrocenter gyroaction obtained in the parallel-symplectic representation is identical to the one obtained in the Hamiltonian representation \cite{Brizard_1989,Brizard_Hahm_2007}.  Next, in Sec.~\ref{sec:3}, we use the gyrocenter phase-space transformation to derive the gyrocenter Hamiltonian up to second order in the perturbation parameter 
$\epsilon$. The gyrocenter Hamilton equations of motion, which are then expressed in terms of the gyrocenter Hamiltonian and the gyrocenter Poisson bracket, are shown to satisfy the gyrocenter Liouville Theorem. In Sec.~\ref{sec:4}, we introduce the gyrokinetic variational principle from which we derive the gyrokinetic Vlasov-Maxwell equations in the parallel-symplectic representation in Sec.~\ref{sec:5}. In addition, we show in Sec.~\ref{sec:5} that the gyrokinetic Vlasov-Maxwell equations possess an exact energy conservation law derived from the gyrokinetic Noether equation derived in Sec.~\ref{sec:4}. We summarize our work in Sec.~\ref{sec:6}, where we also comment on the the truncated-$f$ version of the nonlinear gyrokinetic Vlasov-Maxwell equations  in the parallel-symplectic representation. Lastly, we present two appendices containing detailed derivations of results presented in the text.

\section{\label{sec:2}Derivation of the Gyrocenter Poisson Bracket by Lie-transform Perturbation Method}

In the parallel-symplectic representation presented here, we use the Lie-transform perturbation method to derive the gyrocenter symplectic one-form from the perturbed guiding-center symplectic one-form
\cite{Brizard_1989,Brizard_PhD}:
\begin{equation}
\Gamma_{\rm gy} \;=\; {\sf T}_{\rm gy}^{-1}\left( \Gamma_{0{\rm gc}} \;+\frac{}{} \epsilon\,\Gamma_{1{\rm gc}}\right) \;+\; \exd S \;\equiv\; \Gamma_{0{\rm gy}} + \epsilon\;\Gamma_{1{\rm gy}},
\label{eq:Gamma_gy}
\end{equation}
where the gyrocenter push-forward ${\sf T}_{\rm gy}^{-1} \equiv \cdots \exp(-\epsilon^{2}\pounds_{2{\rm gy}})\,\exp(-\epsilon\,\pounds_{1{\rm gy}})$ is expressed in terms of Lie derivatives 
 $(\pounds_{1{\rm gy}}, \pounds_{2{\rm gy}},\cdots)$ generated by the phase-space vector fields $({\sf G}_{1}, {\sf G}_{2}, \cdots)$ associated with the phase-space transformation to the extended gyrocenter coordinates $\ov{\cal Z}^{a}$, and the gyroangle-dependent gauge functions $(S_{1}, S_{2}, \cdots)$ generate the canonical part of the gyrocenter phase-space transformation. In Eq.~\eqref{eq:Gamma_gy}, the unperturbed gyrocenter one-form is expressed in terms of the unperturbed guiding-center one-form:
\begin{equation}
 \Gamma_{0{\rm gy}} \;\equiv\; \ov{\Gamma}_{0{\rm gc}} \;=\; \left( \frac{e}{c}{\bf A}_{0} + \ov{p}_{\|}\bhat_{0} - \ov{J}\,{\bf R}_{0}^{*} \right)\bdot\exd\ov{\bf X} + \ov{J}\,\exd\ov{\zeta} - \ov{w}\;\exd t,
 \label{eq:Gamma_gc_0}
 \end{equation}
 where the guiding-center gyrogauge vector is ${\bf R}_{0}^{*} \equiv {\bf R} + \frac{1}{2}\,\ov{\nabla}\btimes\bhat_{0}$ (see Eqs.~(13) and (69) of Ref.~\cite{Tronko_Brizard_2015}). 
 
 The first-order perturbed guiding-center one-form in Eq.~\eqref{eq:Gamma_gy} is defined in terms of the guiding-center push-forward of the perturbed particle one-form:
 \begin{equation}
\Gamma_{1{\rm gc}} = {\sf T}_{\rm gc}^{-1}\left( \frac{e}{c}\,{\bf A}_{1}\bdot\exd{\bf x}\right) \;\equiv\; \frac{e}{c}\;{\bf A}_{1{\rm gc}}\bdot\left( \exd{\bf X} + \exd\vb{\rho}_{\rm gc} \right),
 \label{eq:Gamma_gc_1}
 \end{equation}
 where the guiding-center push-forward operator ${\sf T}_{\rm gc}^{-1}$ transforms the perturbed particle one-form into guiding-center phase space, with ${\bf A}_{1{\rm gc}} \equiv 
 {\bf A}_{1}({\bf X} + \vb{\rho}_{\rm gc},t)$. Here, the guiding-center displacement $\vb{\rho}_{\rm gc} \equiv {\sf T}_{\rm gc}^{-1}{\bf x} - {\bf X}$ is defined in terms of the guiding-center push-forward of the particle position ${\bf x}$ (viewed as a function on particle phase space), and the exterior derivative $\exd\vb{\rho}_{\rm gc} \equiv \exd Z^{\alpha}\;(\partial\vb{\rho}_{\rm gc}/\partial Z^{\alpha})$ in Eq.~\eqref{eq:Gamma_gc_1} does not include derivatives with respect to $(w,t)$. 
 
 The purpose of the gyrocenter phase-space transformation is to remove the gyroangle dependence from the first-order gyrocenter symplectic one-form \eqref{eq:Gamma_gc_1}, which, in the parallel-symplectic representation, is transformed into the first-order gyrocenter symplectic one-form
\begin{equation}
\Gamma_{1{\rm gy}} \;\equiv\; \frac{e}{c}\,\langle \ov{A}_{1\|{\rm gc}}\rangle\;\bhat_{0}\bdot\exd\ov{\bf X},
\label{eq:Gamma_gy_1}
\end{equation}
where $\langle \ov{A}_{1\|{\rm gc}}\rangle \equiv \int_{0}^{2\pi}\ov{A}_{1\|{\rm gc}}\,d\ov{\zeta}/2\pi$ denotes the gyroangle-averaged part of $\ov{A}_{1\|{\rm gc}} \equiv A_{1\|}(\ov{\bf X} + \ov{\vb{\rho}}_{0},t)$. Here, and in what follows, we have replaced the full guiding-center displacement $\vb{\rho}_{\rm gc} \rightarrow \vb{\rho}_{0}$ with the lowest-order guiding-center gyroradius $\vb{\rho}_{0}$, which is explicitly gyroangle-dependent. (Higher-order guiding-center corrections, which are not included in the present work, involve the spatial nonuniformity of the background magnetic field ${\bf B}_{0}$ and are discussed in Refs.~\cite{Brizard_2013,Tronko_Brizard_2015}.)

\subsection{First-order gyrocenter transformation}

By substituting Eqs.~\eqref{eq:Gamma_gc_1}-\eqref{eq:Gamma_gy_1} into the first-order terms in Eq.~\eqref{eq:Gamma_gy}, we obtain 
\[ \frac{e}{c}\,\langle \ov{A}_{1\|{\rm gc}}\rangle\;\bhat_{0}\bdot\exd\ov{\bf X}  \;=\; \frac{e}{c}\;\ov{\bf A}_{1{\rm gc}}\bdot\left( \exd\ov{\bf X} \;+\frac{}{} \exd\ov{\vb{\rho}}_{0} \right) \;-\; {\sf G}_{1}\cdot\omega_{0{\rm gc}} \;+\; \exd S_{1},  \] 
which yields the first-order generating-field components $G_{1}^{a}$ as functions of $S_{1}$ and $\ov{\bf A}_{1{\rm gc}}$:
 \begin{eqnarray}
 G_{1}^{a} & = & \{ S_{1},\; \ov{\cal Z}^{a}\}_{0} \;+\; \frac{e}{c}\;\ov{\bf A}_{1{\rm gc}}\bdot\left\{ \ov{\bf X} + \ov{\vb{\rho}}_{0},\frac{}{} \ov{\cal Z}^{a} \right\}_{0} \;-\; \frac{e}{c}\;\langle \ov{A}_{1\|{\rm gc}}\rangle\;\bhat_{0}\bdot\{ \ov{\bf X},\; \ov{\cal Z}^{a}\}_{0} \nonumber \\
  & = & \{ S_{1},\; \ov{\cal Z}^{a}\}_{0} \;+\; \frac{e}{c}\;\ov{\bf A}_{1\bot{\rm gc}}\bdot\left\{ \ov{\bf X} + \ov{\vb{\rho}}_{0},\frac{}{} \ov{\cal Z}^{a} \right\}_{0} \;+\; \frac{e}{c}\;\wt{A}_{1\|{\rm gc}}\;\bhat_{0}\bdot
  \{ \ov{\bf X},\; \ov{\cal Z}^{a}\}_{0},
  \label{eq:G1_alpha}
  \end{eqnarray}
 where $\wt{A}_{1\|{\rm gc}} \equiv \ov{A}_{1\|{\rm gc}} - \langle  \ov{A}_{1\|{\rm gc}}\rangle$ denotes the gyroangle-dependent part of $\ov{A}_{1\|{\rm gc}}$, the unperturbed guiding-center Poisson bracket 
 $\{\;,\;\}_{0}$ is derived from the inversion of the unperturbed gyrocenter Lagrange two-form $\omega_{0{\rm gy}} \equiv \exd\Gamma_{0{\rm gy}}$ obtained from Eq.~\eqref{eq:Gamma_gc_0} [i.e., it is obtained from Eq.~\eqref{eq:PB_gy} in the limit $\epsilon \rightarrow 0$], and we have used $\bhat_{0}\bdot\{\vb{\rho}_{0}, \ov{\cal Z}^{a}\}_{0} = 0$ and we find $\bhat_{0}\bdot\{ \ov{\bf X},\; \ov{\cal Z}^{a}\}_{0} = \partial
\ov{\cal Z}^{a}/\partial\ov{p}_{\|}$. We note that, since the gyrocenter gyroaction $\ov{J}$ satisfies the Darboux constraint 
 $\{ \ov{J},\; \ov{\cal Z}^{a}\}_{\rm gy} \equiv -\;\partial\ov{\cal Z}^{a}/\partial\ov{\zeta}$ on the gyrocenter Poisson bracket, the first-order component
\begin{equation}
G_{1}^{J} \;=\; \pd{S_{1}}{\ov{\zeta}} \;+\; \frac{e}{c}\pd{\ov{\vb{\rho}}_{0}}{\ov{\zeta}}\bdot\ov{\bf A}_{1\bot{\rm gc}}
\label{eq:G1_J}
\end{equation}
is independent of the representation used in gyrokinetic theory since the gauge function $S_{1}$ is also a representation-invariant function [see Eq.~\eqref{eq:S1_dot}]. A simpler way to explain this equivalence of gyrokinetic representations is based on the fact that the component \eqref{eq:G1_J} is identical to the one obtained in the Hamiltonian representation (see Eq.~(166) in Ref.~\cite{Brizard_Hahm_2007}, where the gyrocenter magnetic moment $\ov{\mu} \equiv \ov{J}\,e/mc$ is used instead of the gyrocenter gyroaction $\ov{J}$).

\subsection{Second-order gyrocenter transformation}

At second order $\epsilon^{2}$ in Eq.~\eqref{eq:Gamma_gy}, with $\Gamma_{2{\rm gy}} = 0 = \Gamma_{2{\rm gc}}$, we obtain  
\[ 0 \;=\; -\; {\sf G}_{2}\cdot\omega_{0{\rm gc}} - \frac{1}{2}\,{\sf G}_{1}\cdot(\omega_{1{\rm gc}} + \omega_{1{\rm gy}}) + \exd S_{2} \]
 which yields the second-order generating-field components $G_{2}^{a}$ as functions of $(S_{1}, S_{2})$ and $\ov{\bf A}_{1{\rm gc}}$:
 \begin{eqnarray}
 G_{2}^{a} & = & \{ S_{2},\; \ov{\cal Z}^{a}\}_{0} \;+\; \frac{e}{2c}\left( \ov{\vb{\rho}}_{1{\rm gy}}\bdot\left[\left\{\ov{\bf X} + \ov{\vb{\rho}}_{0},\frac{}{} \ov{\cal Z}^{a}\right\}_{0}\btimes
 \ov{\bf B}_{1{\rm gc}} \;-\; \{ t,\; \ov{\cal Z}^{a}\}_{0}\;\pd{\ov{\bf A}_{1{\rm gc}}}{t}\right] \;-\; G_{1}^{J}\pd{\langle \ov{A}_{1\|{\rm gc}}\rangle}{\ov{J}}\;\bhat_{0}\bdot\{\ov{\bf X},\; \ov{\cal Z}^{a}\}_{0}\right)
\nonumber \\
  &  &-\; \frac{e}{2c}\;G_{1}^{\bf X}\bdot\left[\{\ov{\bf X}, \ov{\cal Z}^{a}\}_{0}\btimes\left(\nabla\btimes(\langle \ov{A}_{1\|{\rm gc}}\rangle \frac{}{} \bhat_{0}) \right) \;-\; \bhat_{0} \left( \{ \ov{J},\; \ov{\cal Z}^{a}\}_{0}
  \pd{\langle \ov{A}_{1\|{\rm gc}}\rangle}{\ov{J}} \;+\;  \{ t,\; \ov{\cal Z}^{a}\}_{0}\;\pd{\langle \ov{A}_{1\|{\rm gc}}\rangle}{t} \right) \right],
   \label{eq:G2_alpha} 
  \end{eqnarray}
where we have used the fact that $G_{1}^{t} \equiv 0$ (in fact, since time is not affected by the gyrocenter phase-space transformation, we have $G_{n}^{t} \equiv 0$ for $n \geq 1$) and we have introduced the guiding-center first-order magnetic field ${\bf B}_{1{\rm gc}} \equiv \nabla\btimes{\bf A}_{1{\rm gc}}$. In addition, the first-order gyrocenter displacement $\ov{\vb{\rho}}_{1{\rm gy}}$ \cite{Brizard_2008,Brizard_2013} is defined as
\begin{eqnarray}
\vb{\rho}_{1{\rm gy}} & \equiv & \left(\frac{d}{d\epsilon}{\sf T}_{\rm gy}^{-1}({\bf X} + \vb{\rho}_{0})\right)_{\epsilon = 0} \;=\; -\; \left(G_{1}^{\bf X} \;+\frac{}{} {\sf G}_{1}\cdot\exd\ov{\vb{\rho}}_{0}\right) \;=\;
\left\{ \ov{\bf X} + \ov{\vb{\rho}}_{0},\frac{}{} S_{1}\right\}_{0}, 
 \label{eq:rho1_gy}
\end{eqnarray}
where we used the guiding-center identities $\{ \ov{\bf X} + \ov{\vb{\rho}}_{0}, \ov{\bf X} + \ov{\vb{\rho}}_{0}\}_{0} = 0 = \bhat_{0}\bdot\{ \ov{\bf X}, \ov{\bf X} + \ov{\vb{\rho}}_{0}\}_{0}$. Lastly, we note that, using the Darboux constraint $\{ \ov{J},\; \ov{\cal Z}^{a}\}_{0} \equiv -\;\partial\ov{\cal Z}^{a}/\partial\ov{\zeta}$ in Eq.~\eqref{eq:G2_alpha}, we obtain the second-order component
\begin{equation}
G_{2}^{J} \;=\; \pd{S_{2}}{\ov{\zeta}} \;-\; \frac{e}{2c}\pd{\ov{\vb{\rho}}_{0}}{\ov{\zeta}}\bdot\left(\ov{\vb{\rho}}_{1{\rm gy}}\btimes\frac{}{}\ov{\bf B}_{1{\rm gc}}\right),
\label{eq:G2_J}
\end{equation}
which is also independent of the representation used in gyrokinetic theory. 

\subsection{\label{subsec:gy_PB}Gyrocenter Poisson bracket}

In the parallel-symplectic representation considered in the present paper, we combine the one-forms \eqref{eq:Gamma_gc_0} and \eqref{eq:Gamma_gy_1} to obtain the extended  gyrocenter phase-space symplectic one-form
\begin{eqnarray}
\Gamma_{\rm gy} & = & \left[ \frac{e}{c} \left( {\bf A}_{0} \;+\frac{}{} \epsilon\,\langle \ov{A}_{1\|{\rm gc}}\rangle\;\bhat_{0}\right) \;+\; \ov{p}_{\|}\;\bhat_{0} \right]\bdot\exd\ov{\bf X} \;+\; \ov{J}\;\left(\exd\ov{\zeta} \;-\frac{}{} 
{\bf R}_{0}^{*}\bdot\exd\ov{\bf X}\right) \;-\; \ov{w}\;\exd t,
\end{eqnarray}
which is then used to derive the gyrocenter Poisson bracket $\{\;,\;\}_{\rm gy}$ as follows \cite{Brizard_Hahm_2007}. First, we compute the gyrocenter two-form $\omega_{\rm gy} \equiv \exd\Gamma_{\rm gy} = \frac{1}{2}\,(\vb{\omega}_{\rm gy})_{ab}\;\exd\ov{\cal Z}^{a}\wedge\exd\ov{\cal Z}^{b}$, whose components form the gyrocenter Lagrange $8\times 8$ antisymmetric matrix $\vb{\omega}_{\rm gy}$. Next, provided the matrix $\vb{\omega}_{\rm gy}$ has a nonvanishing determinant (i.e., ${\rm det}\,\vb{\omega}_{\rm gy} = {\cal J}_{\rm gy}^{2} \neq 0$, where ${\cal J}_{\rm gy}$ denotes the gyrocenter Jacobian), we invert the gyrocenter matrix $\vb{\omega}_{\rm gy}$ to obtain the antisymmetric gyrocenter Poisson matrix whose components define the fundamental gyrocenter Poisson brackets: $J_{\rm gy}^{ab} \equiv 
\{ \ov{\cal Z}^{a},\; \ov{\cal Z}^{b}\}_{\rm gy}$. 

Hence, using this inversion procedure, we construct the gyrocenter Poisson bracket
\begin{eqnarray}
\{ \ov{\cal F},\; \ov{\cal G}\}_{\rm gy} \;\equiv\; \pd{\ov{\cal F}}{\ov{\cal Z}^{a}}\;J_{\rm gy}^{ab}\;\pd{\ov{\cal G}}{\ov{\cal Z}^{b}} & = & \left( \pd{\ov{\cal F}}{\ov{\zeta}}\;\pd{\ov{\cal G}}{\ov{J}} \;-\; \pd{\ov{\cal F}}{\ov{J}}\;\pd{\ov{\cal G}}{\ov{\zeta}} \right) \;+\; \frac{{\bf B}_{\epsilon}^{*}}{B_{\epsilon\|}^{*}}\bdot
\left(\ov{\nabla}_{\epsilon}^{*}\ov{\cal F}\;\pd{\ov{\cal G}}{\ov{p}_{\|}} \;-\; \pd{\ov{\cal F}}{\ov{p}_{\|}}\;\ov{\nabla}_{\epsilon}^{*}\ov{\cal G} \right) \nonumber \\
  &  &-\; \frac{c\,\bhat_{0}}{eB_{\epsilon\|}^{*}}\bdot\left(\ov{\nabla}_{0}^{*}\ov{\cal F}\frac{}{}\btimes\frac{}{}\ov{\nabla}_{0}^{*}\ov{\cal G}\right) \;+\; \left(\pd{\ov{\cal F}}{\ov{w}}\;\pd{\ov{\cal G}}{t} \;-\; 
  \pd{\ov{\cal F}}{t}\;\pd{\ov{\cal G}}{\ov{w}} \right),
\label{eq:PB_gy}
\end{eqnarray}
where the gyrocenter Jacobian 
\begin{equation}
{\cal J}_{\rm gy} \;=\; \frac{e}{c}\,B_{\epsilon\|}^{*} \;=\; \bhat_{0}\bdot \frac{e}{c}\,{\bf B}_{\epsilon}^{*} \;=\; \frac{e}{c}\,B_{0\|}^{*} \;+\; \epsilon\;\frac{e}{c}\,\langle \ov{A}_{1\|{\rm gc}}\rangle\;\bhat_{0}\bdot\ov{\nabla}\btimes\bhat_{0}, 
\label{eq:Jac_gy}
\end{equation}
is defined in terms of the gyrocenter symplectic magnetic field
\begin{equation}
{\bf B}_{\epsilon}^{*} \;\equiv\; \ov{\nabla}\btimes{\bf A}_{\epsilon}^{*} \;=\; \ov{\nabla}\btimes\left({\bf A}_{0}^{*} \;+\; \epsilon\,\langle \ov{A}_{1\|{\rm gc}}\rangle\;\bhat_{0} \right) \;=\; {\bf B}_{0}^{*} \;+\; \epsilon\;\ov{\nabla}\btimes\left(\langle \ov{A}_{1\|{\rm gc}}\rangle\;\bhat_{0}\right), 
\label{eq:B_star} 
\end{equation}
with ${\bf A}_{0}^{*} \equiv {\bf A}_{0} + (c/e)\;(\ov{p}_{\|}\,\bhat_{0} - \ov{J}\,{\bf R}_{0}^{*})$ and $B_{0\|}^{*} \equiv \bhat_{0}\bdot{\bf B}_{0}^{*}$. In the gyrocenter Poisson bracket 
\eqref{eq:PB_gy}, the modified gradient operator $\nabla_{\epsilon}^{*}$ is defined as
\begin{equation}
\ov{\nabla}_{\epsilon}^{*} \equiv \ov{\nabla}_{0}^{*} \;-\; \epsilon\,\frac{e\bhat_{0}}{c} \left( \pd{\langle \ov{A}_{1\|{\rm gc}}\rangle}{t}\;\pd{}{\ov{w}} + \pd{\langle \ov{A}_{1\|{\rm gc}}\rangle}{\ov{J}}\;\pd{}{\ov{\zeta}} \right),
\end{equation}
where $\ov{\nabla}_{0}^{*} \equiv \ov{\nabla} + {\bf R}_{0}^{*}\,\partial/\partial\ov{\zeta}$. We note that the guiding-center Poisson bracket $\{\;,\;\}_{\rm gc} \equiv \{\;,\;\}_{0}$ used in the Hamiltonian representation is obtained from the gyrocenter Poisson bracket \eqref{eq:PB_gy} by setting the perturbation parameter $\epsilon = 0$.

Lastly, we note that the gyrocenter Poisson bracket \eqref{eq:PB_gy} satisfies the standard properties of a Poisson bracket (i.e., antisymmetry, Leibniz, and Jacobi), where the Jacobi identity 
\begin{equation}
\left\{ \ov{\cal F} ,\frac{}{} \{ \ov{\cal G},\; \ov{\cal H}\}_{\rm gy} \right\}_{\rm gy} \;+\; \left\{ \ov{\cal G},\frac{}{} \{ \ov{\cal H},\; \ov{\cal F}\}_{\rm gy} \right\}_{\rm gy} \;+\; \left\{ \ov{\cal H},\frac{}{} \{ \ov{\cal F},\; 
\ov{\cal G}\}_{\rm gy} \right\}_{\rm gy} \;=\; 0,
\label{eq:Jac_gy}
\end{equation}
which holds for any three functions $(\ov{\cal F},\ov{\cal G},\ov{\cal H})$ on extended gyrocenter phase space, follows from the fact that the gyrocenter symplectic magnetic field \eqref{eq:B_star} is divergenceless: 
$\ov{\nabla}\bdot{\bf B}_{\epsilon}^{*} \equiv \ov{\nabla}\bdot(\ov{\nabla}\btimes{\bf A}_{\epsilon}^{*}) = 0$. In addition, we show in App.~\ref{sec:App_A} that the gyrocenter Poisson bracket \eqref{eq:PB_gy} satisfies the Liouville property
\begin{equation}
\{ \ov{\cal F},\; \ov{\cal G}\}_{\rm gy} \;\equiv\; \frac{1}{B_{\epsilon \|}^{*}}\;\pd{}{\ov{\cal Z}^{a}}\left( B_{\epsilon \|}^{*}\frac{}{} \{ \ov{\cal F},\; \ov{\cal Z}^{a}\}_{\rm gy}\;\ov{\cal G} \right),
\label{eq:Liouville_gyPB}
\end{equation}
which hold for all functions $(\ov{\cal F}, \ov{\cal G})$ and for any value of the perturbation ordering parameter $\epsilon$ (including $\epsilon = 0$). This property guarantees that the identity $\int {\cal J}_{\rm gy}\,
\{ \ov{\cal F},\; \ov{\cal G}\}_{\rm gy}  d^{8}\ov{\cal Z} \equiv 0$ is always satisfied .

\section{\label{sec:3}Gyrocenter Hamiltonian in the Parallel-Symplectic Representation}

By using the gyrocenter Poisson bracket \eqref{eq:PB_gy}, the gyrocenter Hamilton equations in extended gyrocenter phase space are expressed as $\dot{\ov{\cal Z}}^{a}_{\rm gy} \equiv \{ \ov{\cal Z}^{a},
{\cal H}_{\rm gy}\}_{\rm gy}$ in terms of the extended gyrocenter Hamiltonian ${\cal H}_{\rm gy} \equiv H_{\rm gy} - w$. In particular, the gyrocenter Hamilton equations for $\ov{\bf X}$ and $\ov{p}_{\|}$ are
\begin{eqnarray}
\dot{\ov{\bf X}}_{\rm gy} & \equiv & \{ \ov{\bf X},\; {\cal H}_{\rm gy}\}_{\rm gy} \;=\; \pd{H_{\rm gy}}{\ov{p}_{\|}}\;\frac{{\bf B}_{\epsilon}^{*}}{B_{\epsilon\|}^{*}} \;+\; 
\frac{c\bhat_{0}}{eB_{\epsilon\|}^{*}}\btimes\ov{\nabla} H_{\rm gy}, 
\label{eq:xdot_gy} \\
\dot{\ov{p}}_{\|{\rm gy}} & \equiv & \{ \ov{p}_{\|},\; {\cal H}_{\rm gy}\}_{\rm gy} \;=\; -\;\frac{{\bf B}_{\epsilon}^{*}}{B_{\epsilon\|}^{*}}\bdot\ov{\nabla} H_{\rm gy} \;-\; \epsilon\,\frac{e}{c}
\pd{\langle \ov{A}_{1\|{\rm gc}}\rangle}{t},
\label{eq:pdot_gy}
\end{eqnarray}
where we note that the gyrocenter parallel velocity
\begin{equation}
\bhat_{0}\bdot\dot{\ov{\bf X}}_{\rm gy} \;\equiv\; \pd{H_{\rm gy}}{\ov{p}_{\|}} \;=\; \frac{\ov{p}_{\|}}{m} \;+\; \epsilon^{2}\;\pd{H_{2{\rm gy}}}{\ov{p}_{\|}} 
\label{eq:xdot_par}
\end{equation}
includes the unperturbed component $\ov{p}_{\|}/m$ and a second-order correction (we will show below that $\partial H_{1{\rm gy}}/\partial \ov{p}_{\|} \equiv 0$ in the parallel-symplectic representation). In addition, within the gyrocenter Hamiltonian dynamics, we note that the gyrocenter gyroaction $\ov{J}$ is a gyrocenter invariant (i.e., $\dot{\ov{J}}_{\rm gy} = -\,\partial H_{\rm gy}/\partial\ov{\zeta} \equiv 0$). We remark that the gyrocenter parallel-force equation \eqref{eq:pdot_gy} includes a contribution from the parallel inductive electric field in the parallel-symplectic representation, which is absent in the Hamiltonian representation. 

We also note that the gyrocenter Hamilton equations \eqref{eq:xdot_gy}-\eqref{eq:pdot_gy} satisfy the gyrocenter Liouville Theorem
\begin{equation}
\pd{{\cal J}_{\rm gy}}{t} \;+\; \ov{\nabla}\bdot\left({\cal J}_{\rm gy}\frac{}{} \dot{\ov{\bf X}}_{\rm gy}\right) \;+\; \pd{}{\ov{p}_{\|}}\left({\cal J}_{\rm gy}\frac{}{} \dot{\ov{p}}_{\|{\rm gy}}\right) \;=\; 0.
\label{eq:gy_Liouville}
\end{equation}
The proof of Eq.~\eqref{eq:gy_Liouville} follows simply from the expressions
\begin{eqnarray*}
\ov{\nabla}\bdot\left({\cal J}_{\rm gy}\frac{}{} \dot{\ov{\bf X}}_{\rm gy}\right) & = & \frac{e}{c}\,{\bf B}_{\epsilon}^{*}\bdot\ov{\nabla}\pd{H_{\rm gy}}{\ov{p}_{\|}} \;+\; \frac{e}{c}\,\pd{{\bf B}_{\epsilon}^{*}}{\ov{p}_{\|}}\bdot\ov{\nabla}H_{\rm gy}, \\
\pd{}{\ov{p}_{\|}}\left({\cal J}_{\rm gy}\frac{}{} \dot{\ov{p}}_{\|{\rm gy}}\right) & = & -\;\frac{e}{c}\,{\bf B}_{\epsilon}^{*}\bdot\ov{\nabla}\pd{H_{\rm gy}}{\ov{p}_{\|}} \;-\; \frac{e}{c}\,\pd{{\bf B}_{\epsilon}^{*}}{\ov{p}_{\|}}\bdot\ov{\nabla}H_{\rm gy} \;-\; \epsilon\,\frac{e}{c}\,\pd{\langle\ov{A}_{1\|{\rm gc}}\rangle}{t}\;\pd{{\cal J}_{\rm gy}}{\ov{p}_{\|}},
\end{eqnarray*}
where we used $\ov{\nabla}\bdot{\bf B}_{\epsilon}^{*} = 0$, $\partial{\bf B}_{\epsilon}^{*}/\partial\ov{p}_{\|} = \ov{\nabla}\btimes(c\bhat_{0}/e)$. We now recover the gyrocenter Liouville Theorem
\eqref{eq:gy_Liouville} provided the gyrocenter Liouville condition
\begin{equation} 
\epsilon\,\frac{e}{c}\,\pd{\langle\ov{A}_{1\|{\rm gc}}\rangle}{t}\;\pd{{\cal J}_{\rm gy}}{\ov{p}_{\|}}  \;=\; \pd{{\cal J}_{\rm gy}}{t} 
\label{eq:Jac_gy_dot}
\end{equation}
is satisfied, which requires that the first-order correction is kept in the gyrocenter Jacobian \eqref{eq:Jac_gy}. In addition to guaranteeing the proper Hamiltonian dynamics in gyrocenter phase space, we will show in Sec.~\ref{sec:5} how the gyrocenter Liouville condition \eqref{eq:Jac_gy_dot} also guarantees energy conservation.

Up to now, the gyrocenter Hamilton equations \eqref{eq:xdot_gy}-\eqref{eq:pdot_gy} and the gyrocenter Liouville Theorem \eqref{eq:gy_Liouville} are valid for any Hamiltonian that satisfies the gyroaction-invariance property $\dot{\ov{J}}_{\rm gy} = -\,\partial H_{\rm gy}/\partial\ov{\zeta} \equiv 0$. We now proceed with the derivation of the gyrocenter Hamiltonian 
\begin{equation}
{\cal H}_{\rm gy} \equiv {\sf T}_{\rm gy}^{-1}{\cal H}_{\rm gc} = {\cal H}_{0{\rm gc}} + \epsilon \left( e\,\Phi_{1{\rm gc}} - {\sf G}_{1}\cdot\exd{\cal H}_{0{\rm gc}} \right) + \epsilon^{2} \left[ \frac{1}{2}\,
{\sf G}_{1}\cdot\exd\left({\sf G}_{1}\cdot\exd{\cal H}_{0{\rm gc}}\right) - {\sf G}_{1}\cdot\exd (e\,\Phi_{1{\rm gc}}) - {\sf G}_{2}\cdot\exd{\cal H}_{0{\rm gc}} \right] + \cdots
\end{equation}
by Lie-transform perturbation method by using the components \eqref{eq:G1_alpha} and \eqref{eq:G2_alpha} of the phase-space generating vector fields ${\sf G}_{1}$ and ${\sf G}_{2}$ associated with the gyrocenter transformation. Here, the lowest-order extended guiding-center Hamiltonian is ${\cal H}_{0{\rm gc}} \equiv K_{\rm gc} - w$, where $K_{\rm gc} = J\,\Omega_{0} + p_{\|}^{2}/2m$ denotes the unperturbed guiding-center kinetic energy.

\subsection{First-order gyrocenter Hamiltonian}

The gyroangle-independent part of the first-order gyrocenter Hamiltonian equation 
\begin{equation}
H_{1{\rm gy}} \;=\; e\,\ov{\Phi}_{1{\rm gc}} \;-\; G_{1}^{a}\,\partial_{a}{\cal H}_{0{\rm gc}} \;=\; e\,\ov{\Phi}_{1{\rm gc}} \;-\; \frac{e\Omega}{c}\;\ov{\bf A}_{1\bot{\rm gc}}\bdot\pd{\ov{\vb{\rho}}_{0}}{\ov{\zeta}} \;-\; \left( 
\left\{ S_{1},\frac{}{} \ov{\cal H}_{0{\rm gc}}\right\}_{0} \;+\; \frac{e\ov{p}_{\|}}{mc}\;\wt{A}_{1\|{\rm gc}}\right),
\label{eq:H1gy}
\end{equation}
yields the first-order gyrocenter Hamiltonian
 \begin{eqnarray}
 H_{1{\rm gy}}  &= & e\,\langle\ov{\Phi}_{1{\rm gc}}\rangle \;-\; \frac{e\Omega}{c}\,\left\langle \ov{\bf A}_{1\bot{\rm gc}}\bdot\pd{\ov{\vb{\rho}}_{0}}{\ov{\zeta}}\right\rangle \;\equiv\; e\,\langle\ov{\chi}_{1{\rm gc}}\rangle,
 \label{eq:H1_gy}
  \end{eqnarray}
where the first-order components $G_{1}^{a}$ are defined in Eq.~\eqref{eq:G1_alpha}.  Here, the parallel component $\ov{A}_{1\|}$ is absent from Eq.~\eqref{eq:H1_gy}, and the first-order gyrocenter Hamiltonian \eqref{eq:H1_gy} does not depend on the parallel kinetic momentum $\ov{p}_{\|}$ (i.e., $\partial H_{1{\rm gy}}/\partial \ov{p}_{\|} = 0$), so that Eq.~\eqref{eq:xdot_par} yields
$\partial H_{\rm gy}/\partial\ov{p}_{\|} - \ov{p}_{\|}/m \equiv {\cal O}(\epsilon^{2})$. 

The gyroangle-dependent part of Eq.~\eqref{eq:H1gy} yields the Lie-transform equation for the first-order gauge function $S_{1}$:
 \begin{equation}
 \frac{d_{\rm gc}S_{1}}{dt} \;\equiv\; \{ S_{1},\; \ov{\cal H}_{0{\rm gc}}\}_{0} \;=\; e\;\wt{\Phi}_{1{\rm gc}} \;-\; \frac{e\ov{p}_{\|}}{mc}\;\wt{A}_{1\|{\rm gc}} \;-\; \frac{e\Omega}{c} \left( \ov{\bf A}_{1\bot{\rm gc}}\bdot\pd{\ov{\vb{\rho}}_{0}}{\ov{\zeta}} \;-\; \left\langle \ov{\bf A}_{1\bot{\rm gc}}\bdot\pd{\ov{\vb{\rho}}_{0}}{\ov{\zeta}}\right\rangle\right) \;\equiv\; e\,\wt{\psi}_{1{\rm gc}}.
 \label{eq:S1_dot}
 \end{equation}
We note that the definition of the first-order gauge function $S_{1}$ is independent of the representation used in gyrokinetic theory. Indeed, to lowest order in the guiding-center ordering and finite-Larmor-radius (FLR) effects, where $d_{\rm gc}S_{1}/dt \simeq \Omega_{0}\,\partial S_{1}/\partial\ov{\zeta}$, the first-order gauge function is defined by the indefinite gyroangle integral
\begin{equation} 
S_{1} \;\simeq\; \frac{e}{\Omega_{0}}\int \wt{\psi}_{1{\rm gc}}\;d\ov{\zeta} \;\simeq\; -\,m\,\ov{\vb{\rho}}_{0}\bdot\left(\frac{c\bhat_{0}}{B_{0}}\btimes\ov{\nabla}\ov{\Phi}_{1} \;+\; \frac{\ov{p}_{\|}}{mB_{0}}\;
\ov{\bf B}_{1\bot} \;+\; \frac{e}{mc}\,\ov{\bf A}_{1\bot}\right), 
\label{eq:S1_approx}
\end{equation}
which is identical to the first-order gyrocenter gauge function obtained in the Hamiltonian representation (see Sec.~3.2 of Ref.~\cite{Brizard_PhD} for more details). Here, we used the approximation
$\ov{\bf B}_{1\bot} \simeq \ov{\nabla}\ov{A}_{1\|}\btimes\bhat_{0}$.
 
 \subsection{Second-order gyrocenter Hamiltonian}
 
The gyroangle-independent part of the second-order gyrocenter Hamiltonian equation 
\begin{eqnarray}
H_{2{\rm gy}} & = & -\, G_{2}^{a}\,\partial_{a}{\cal H}_{0{\rm gc}} \;-\; \frac{1}{2}\,G_{1}^{a}\,\partial_{a}(e\,\Phi_{1{\rm gc}} \;+\; H_{1{\rm gy}}) \nonumber \\
 & = & -\;\frac{e}{2}\,\vb{\rho}_{1{\rm gy}}\bdot\left(\ov{\bf E}_{1{\rm gc}} + \frac{1}{c}\left\{ \ov{\bf X} + \ov{\vb{\rho}}_{0},\frac{}{} \ov{K}_{\rm gc}\right\}_{0}\btimes \ov{\bf B}_{1{\rm gc}}\right) \;-\;
 \frac{e}{2c}\,\ov{\bf A}_{1\bot{\rm gc}}\bdot\left\{ \ov{\bf X} + \ov{\vb{\rho}}_{0},\frac{}{} e\,\langle\ov{\chi}_{1{\rm gc}}\rangle\right\}_{0} \nonumber \\
  &  &+ \frac{e}{2c} \left[ \left( G_{1}^{J}\,\pd{}{\ov{J}} + G_{1}^{\bf X}\bdot\ov{\nabla}\right)\frac{\ov{p}_{\|}}{m}\,\langle A_{1\|{\rm gc}}\rangle  - G_{1}^{\bf X}\bdot\bhat_{0}\;
  \frac{d_{0}\langle A_{1\|{\rm gc}}\rangle}{dt} \right] - \left\{ S_{2},\frac{}{} \ov{\cal H}_{0{\rm gc}} \right\}_{0} - \frac{1}{2}\,\left\{ S_{1},\frac{}{} e\,\langle\ov{\chi}_{1{\rm gc}}\rangle\right\}_{0}
\end{eqnarray} 
yields the second-order gyrocenter Hamiltonian
 \begin{eqnarray}
 H_{2{\rm gy}} & = & -\;\frac{e}{2} \left\langle \vb{\rho}_{1{\rm gy}}\bdot\left(\ov{\bf E}_{1{\rm gc}} + \frac{1}{c}\left\{ \ov{\bf X} + \ov{\vb{\rho}}_{0},\frac{}{} \ov{K}_{\rm gc}
 \right\}_{0}\btimes \ov{\bf B}_{1{\rm gc}}\right)\right\rangle - \frac{e}{2c}\,\left\langle \ov{\bf A}_{1\bot{\rm gc}}\bdot\left\{ \ov{\bf X} + \ov{\vb{\rho}}_{0},\frac{}{} e\,\langle\ov{\psi}_{1{\rm gc}}\rangle\right\}_{0} \right\rangle, 
    \label{eq:H2_gy0}
  \end{eqnarray}
where $\langle\ov{\psi}_{1{\rm gc}}\rangle \equiv \langle\ov{\chi}_{1{\rm gc}}\rangle - \langle\ov{A}_{1\|{\rm gc}}\rangle\,\ov{p}_{\|}/mc$, the second-order components $G_{2}^{a}$ are defined in Eq.~\eqref{eq:G2_alpha}, while the first-order gyrocenter displacement $\vb{\rho}_{1{\rm gy}}$ is defined in Eq.~\eqref{eq:rho1_gy}, and we introduced the guiding-center first-order electric field $\ov{\bf E}_{1{\rm gc}} \equiv -\,\ov{\nabla}\ov{\Phi}_{1{\rm gc}} - c^{-1}\partial\ov{\bf A}_{1{\rm gc}}/\partial t$. We note that the equation for the second-order gauge function $S_{2}$ is not needed in what follows since we are not interested in continuing the derivation of the gyrocenter Hamiltonian beyond the second order.

In App.~\ref{sec:App_B}, we derive the identity
\begin{eqnarray}
e\left\langle \{ S_{1}, \ov{\bf X} + \ov{\vb{\rho}}_{0}\}_{0}\bdot\left(\ov{\bf E}_{1{\rm gc}} + \frac{1}{c}\{ {\bf X} + \ov{\vb{\rho}}_{0}, \ov{K}_{\rm gc}\}_{0}\btimes \ov{\bf B}_{1{\rm gc}}\right)\right\rangle & = & -\;\frac{e}{c}\,\left\langle \ov{\bf A}_{1{\rm gc}}\bdot\left\{ \ov{\bf X} + \ov{\vb{\rho}}_{0},\frac{}{} e\,\wt{\psi}_{1{\rm gc}}\right\}_{0} \right\rangle \nonumber \\
 &  &-\; \left\langle \left\{ S_{1},\frac{}{} e\,\wt{\psi}_{1{\rm gc}}\right\}_{0}\right\rangle \;+\; \frac{e}{c}\,\frac{d_{\rm gc}}{dt}\langle \ov{\bf A}_{1{\rm gc}}\bdot\ov{\vb{\rho}}_{1{\rm gy}}\rangle,
 \label{eq:2gy_id}
\end{eqnarray} 
where we used the first-order equation \eqref{eq:S1_dot} for $S_{1}$. Hence,  the second-order gyrocenter Hamiltonian \eqref{eq:H2_gy0} can also be written as
\begin{equation}
H_{2{\rm gy}} \;=\; \frac{e^{2}}{2mc^{2}} \left(  \left\langle|\ov{\bf A}_{1\bot{\rm gc}}|^{2}\right\rangle \;+\frac{}{} \left\langle (\wt{A}_{1\|{\rm gc}})^{2}\right\rangle \right) \;-\; \frac{1}{2}\;\left\langle \left\{ S_{1},\frac{}{} 
\{ S_{1},\; \ov{\cal H}_{0{\rm gc}}\}_{0}\right\}_{0}\right\rangle \;\equiv\; -\;\frac{1}{2}\left\langle \pounds_{1{\rm gy}}\left(e\frac{}{}\ov{\psi}_{1{\rm gc}}\right)\right\rangle,
\label{eq:H2_gy}
\end{equation}
where we used Eq.~\eqref{eq:S1_dot} and we eliminated the exact time derivative (which can be ignored in Hamiltonian mechanics). We note that the second-order gyrocenter Hamiltonian \eqref{eq:H2_gy} depends on $\ov{p}_{\|}$ only through the gyrocenter ponderomotive Hamiltonian $-\,\frac{1}{2}\langle\{ S_{1},\;\{ S_{1},\; \ov{\cal H}_{0{\rm gc}}\}_{0}\}_{0}\rangle$. Indeed, we find
\begin{equation}
\pd{H_{2{\rm gy}}}{\ov{p}_{\|}} \;=\; -\;\left\langle \left\{ S_{1}, e\,\pd{\wt{\psi}_{1{\rm gc}}}{\ov{p}_{\|}}\right\}_{0}\right\rangle \;=\; \frac{e}{mc}\;\left\langle\{ S_{1},\frac{}{} \wt{A}_{1\|{\rm gc}}\}_{0}\right\rangle \;\simeq\;
-\;\frac{e}{mc}\,\langle\vb{\rho}_{1{\rm gy}}\rangle\bdot\ov{\nabla}A_{1\|},
\label{eq:H2_p}
\end{equation}
which, to lowest order in FLR effects, becomes
\begin{equation}
\pd{H_{2{\rm gy}}}{\ov{p}_{\|}} \;\simeq\; -\;\frac{{\bf B}_{1\bot}}{B_{0}} \bdot\left( \frac{c\bhat_{0}}{B_{0}}\btimes\ov{\nabla}\Phi_{1} \;+\; \frac{\ov{p}_{\|}}{m}\,\frac{{\bf B}_{1\bot}}{B_{0}} \;+\; \frac{e}{mc}\,
{\bf A}_{1\bot} \right),
\end{equation}
where we used Eq.~\eqref{eq:S1_approx}.

\subsection{Gyrocenter Hamiltonian}

By combining the first-order and second-order gyrocenter Hamiltonians \eqref{eq:H1_gy} and \eqref{eq:H2_gy}, the gyrocenter Hamiltonian is defined in the parallel-symplectic representation as
\begin{eqnarray}
H_{\rm gy} & = & \frac{\ov{p}_{\|}^{2}}{2m} \;+\; \ov{J}\,\Omega_{0} \;+\; \epsilon \left( e\,\langle\ov{\Phi}_{1{\rm gc}}\rangle \;-\; \frac{e\Omega}{c}\;\left\langle\ov{\bf A}_{1\bot{\rm gc}}\bdot\pd{\ov{\vb{\rho}}_{0}}{\ov{\zeta}}\right\rangle \right) \;-\; \frac{\epsilon^{2}}{2}\left\langle \pounds_{1{\rm gy}}\left(e\frac{}{}\ov{\psi}_{1{\rm gc}}\right)\right\rangle.
 \label{eq:H_gy}
\end{eqnarray}
From this expression and Eq.~\eqref{eq:H2_p}, we easily find the parallel gyrocenter velocity
\begin{equation}
\bhat_{0}\bdot\dot{\ov{\bf X}}_{\rm gy} \;=\; \pd{H_{\rm gy}}{\ov{p}_{\|}} \;=\; \frac{\ov{p}_{\|}}{m} \;+\; \epsilon^{2}\,\frac{e}{mc}\left\langle\{ S_{1},\frac{}{}\wt{A}_{1\|{\rm gc}}\}_{0}\right\rangle.
\label{eq:H_p||}
\end{equation}
We note that this expression corresponds to the gyroangle-averaged gyrocenter push-forward $\langle{\sf T}_{\rm gy}^{-1}(p_{\|}/m)\rangle$ of the guiding-center parallel kinetic momentum $p_{\|}$ (viewed here as a function on guiding-center phase space), where ${\sf T}_{\rm gy}^{-1}p_{\|}$ is expanded up to second order in the perturbation parameter $\epsilon$: ${\sf T}_{\rm gy}^{-1}p_{\|} = 
\ov{p}_{\|} - \epsilon\,G_{1}^{p_{\|}} - \epsilon^{2}\,(G_{2}^{p_{\|}} - \frac{1}{2}\,{\sf G}_{1}\cdot\exd G_{1}^{p_{\|}})$, with $\langle G_{1}^{p_{\|}}\rangle = \langle \bhat_{0}\bdot\ov{\nabla}S_{1} + (e/c)\,
\wt{A}_{1\|{\rm gc}}\rangle = 0$.

Lastly, the Eulerian variation of the gyrocenter Hamiltonian \eqref{eq:H_gy} appearing in the gyrokinetic variational principle [see Eq.~\eqref{eq:deltaA_gy} below] can be expressed as
 \begin{eqnarray}
\delta H_{\rm gy} & = & \epsilon \left( e\,\langle\delta\ov{\Phi}_{1{\rm gc}}\rangle \;-\; \frac{e\Omega}{c}\;\left\langle\delta\ov{\bf A}_{1\bot{\rm gc}}\bdot\pd{\ov{\vb{\rho}}_{0}}{\ov{\zeta}}\right\rangle \right) \;-\; 
\epsilon^{2}\; \left\langle \pounds_{1{\rm gy}}(e\frac{}{}\delta\ov{\psi}_{1{\rm gc}})\right\rangle \nonumber \\
 & \equiv & \int \left[ \epsilon\,\delta\Phi_{1}({\bf x})\left(\epsilon^{-1}\fd{H_{\rm gy}}{\Phi_{1}({\bf x})}\right) \;+\; \epsilon\,\delta A_{1\|}({\bf x})\left(\epsilon^{-1}\fd{H_{\rm gy}}{A_{1\|}({\bf x})}\right) \;+\; 
 \epsilon\,\delta {\bf A}_{1\bot}({\bf x})\bdot\left(\epsilon^{-1}\fd{H_{\rm gy}}{{\bf A}_{1\bot}({\bf x})} \right)\right] d^{3}x,
 \label{eq:delta_Hgy}
\end{eqnarray}
where the functional derivatives
\begin{eqnarray}
\epsilon^{-1}\fd{H_{\rm gy}}{\Phi_{1}({\bf x})} & = & \langle e\,\delta_{\rm gc}^{3}\rangle \;-\; \epsilon\; \left\langle \pounds_{1{\rm gy}}(e\frac{}{}\delta_{\rm gc}^{3})\right\rangle, \label{eq:fd_phi} \\
\epsilon^{-1}\fd{H_{\rm gy}}{A_{1\|}({\bf x})} & = & \epsilon\; \left\langle \pounds_{1{\rm gy}}\left(e\,\delta_{\rm gc}^{3}\;\frac{\ov{p}_{\|}}{mc}\right)\right\rangle, \label{eq:fd_A_par} \\
\epsilon^{-1}\fd{H_{\rm gy}}{{\bf A}_{1\bot}({\bf x})} & = & -\;\frac{e\Omega}{c}\;\left\langle\delta_{\rm gc}^{3}\pd{\ov{\vb{\rho}}_{0}}{\ov{\zeta}}\right\rangle \;+\; \epsilon\; \left\langle 
\pounds_{1{\rm gy}}\left(\frac{e\Omega}{c}\delta_{\rm gc}^{3}\pd{\ov{\vb{\rho}}_{0}}{\ov{\zeta}}\right)\right\rangle \label{eq:fd_A_perp}
\end{eqnarray}
are expressed in terms of the guiding-center spatial delta function $\delta^{3}_{\rm gc} \equiv \delta^{3}(\ov{\bf X} + \ov{\vb{\rho}}_{0} - {\bf x})$. We note that, by definition, these functional derivatives automatically reduce the maximum perturbation order by one (i.e., a gyrocenter Hamiltonian truncated at second order in $\epsilon$ yields functional derivatives that are truncated at first order in $\epsilon$).

\section{\label{sec:4}Gyrokinetic Variational Principle}

In order to derive a self-consistent set of energy-conserving gyrokinetic Vlasov-Maxwell equations in the parallel-symplectic representation, we now introduce the following gyrokinetic variational principle
\cite{Brizard_2000a,Brizard_2000b}. First, the action functional for the gyrokinetic Vlasov-Maxwell equations in the parallel-symplectic representation is
\begin{equation}
{\cal A}_{\rm gy}[\ov{F}, \Phi_{1}, {\bf A}_{1}] \;=\; -\;\sum \int {\cal F}_{\rm gy}\,{\cal H}_{\rm gy}\;d^{8}\ov{\cal Z} \;+\; \int \frac{d^{3}x\,dt}{8\pi} \left( \epsilon^{2}\;|\nabla\Phi_{1}|^{2} \;-\frac{}{} |{\bf B}_{0} + \epsilon\,\nabla\btimes{\bf A}_{1}|^{2} \right),
\label{eq:A_gy}
\end{equation}
where the variational fields are the gyrocenter Vlasov distribution $\ov{F}$ and the first-order perturbed potentials $(\Phi_{1}, {\bf A}_{1})$, while $\sum$ denotes a sum over particle species and $d^{8}\ov{\cal Z} \equiv d^{6}\ov{Z}\,d\ov{w}\,dt$. We note that the gyrokinetic action functional \eqref{eq:A_gy} has the same form in all representations and all differences between representations will arise from the specific forms of the Eulerian variations $\delta{\cal F}_{\rm gy}$ and $\delta{\cal H}_{\rm gy}$. Here, the extended gyrocenter Vlasov density 
\begin{equation}
{\cal F}_{\rm gy}(\ov{\cal Z}) \;\equiv\; {\cal J}_{\rm gy}\;\ov{F}(\ov{\bf X}, \ov{p}_{\|}, t; \ov{J})\;\delta(\ov{w} - H_{\rm gy}) \;=\; {\cal J}_{\rm gy}\;\ov{\cal F}(\ov{\cal Z})
\label{eq:calF_gy}
\end{equation}
includes the energy delta function $\delta(\ov{w} - H_{\rm gy})$, which ensures that the physical gyrocenter motion in extended gyrocenter phase space takes place on the surface ${\cal H}_{\rm gy} = H_{\rm gy} - \ov{w} \equiv 0$ \cite{Brizard_2000a,Brizard_2000b}, and the time-dependent gyrocenter Jacobian ${\cal J}_{\rm gy} = (e/c)\,B_{\epsilon\|}^{*}$ depends on ${\bf A}_{1}$ through $\langle \ov{A}_{1\|{\rm gc}}\rangle$.
In addition, the gyrocenter Vlasov distribution function $\ov{F} \equiv {\sf T}_{\rm gy}^{-1}F \equiv {\sf T}_{\rm gy}^{-1}({\sf T}_{\rm gc}^{-1}f)$ is defined as the gyrocenter push-forward of the guiding-center Vlasov distribution function $F$, which in turn is defined as the guiding-center push-forward of the particle Vlasov distribution function $f$ \cite{Brizard_1994}. 

\subsection{Eulerian variations}

We now evaluate the Eulerian variation $\delta{\cal A}_{\rm gy}$ of the gyrocenter action functional \eqref{eq:A_gy}:
\begin{equation}
\delta{\cal A}_{\rm gy} \;=\; -\;\sum \int \left( \delta{\cal F}_{\rm gy}\,{\cal H}_{\rm gy} \;+\frac{}{} {\cal F}_{\rm gy}\,\delta H_{\rm gy}\right)d^{8}\ov{\cal Z} \;+\; \int \frac{d^{3}x\,dt}{4\pi} \left( \epsilon^{2}\;
\nabla\delta\Phi_{1}\bdot\nabla\Phi_{1} \;-\frac{}{} \epsilon\,\nabla\btimes\delta{\bf A}_{1}\bdot{\bf B} \right),
\label{eq:deltaA_gy}
\end{equation}
where ${\bf B} = {\bf B}_{0} + \epsilon\,{\bf B}_{1}$ denotes the total magnetic field and the Eulerian variation $\delta H_{\rm gy}$ of the gyrocenter Hamiltonian is given by Eq.~\eqref{eq:delta_Hgy}. The Maxwell variation, on the other hand, is
\[ \frac{\epsilon^{2}}{4\pi}\;\nabla\delta\Phi_{1}\bdot\nabla\Phi_{1}  - \frac{\epsilon}{4\pi}\,\nabla\btimes\delta{\bf A}_{1}\bdot{\bf B} = -\;\epsilon\,\delta\Phi_{1}\;\left(\frac{\epsilon}{4\pi}\nabla^{2}\Phi_{1}\right) -
\epsilon\,\delta{\bf A}_{1}\bdot\left(\frac{\nabla\btimes{\bf B}}{4\pi}\right) + \nabla\bdot\left(\epsilon^{2}\,\frac{\delta\Phi_{1}}{4\pi}\;\nabla\Phi_{1} - \epsilon\frac{\delta{\bf A}_{1}}{4\pi}\btimes{\bf B}\right), \]
which yields
\begin{equation}
-\;\epsilon\,\int \left[ \delta\Phi_{1}\;\left(\frac{\epsilon}{4\pi}\nabla^{2}\Phi_{1}\right) \;-\; \delta{\bf A}_{1}\bdot\left(\frac{\nabla\btimes{\bf B}}{4\pi}\right) \right] d^{3}x,
\label{eq:Maxwell_var}
\end{equation}
where the space divergence terms contribute to the gyrokinetic Noether equation [see Eq.~\eqref{eq:gy_Noether} below] as we will show in the next Section. Here, the surface terms vanish since $\delta\Phi_{1}$ and 
$\delta{\bf A}_{1}$ are assumed to vanish on the integration boundary.

Lastly, we evaluate the constrained Eulerian variation of the extended gyrocenter Vlasov density \eqref{eq:calF_gy}:
\begin{eqnarray}
\delta{\cal F}_{\rm gy} & = & \delta{\cal J}_{\rm gy}\;\ov{\cal F} \;+\; {\cal J}_{\rm gy}\;\delta\ov{\cal F} \nonumber \\
 & = & \left(\epsilon\; \frac{e}{c}\,\langle\delta \ov{A}_{1\|{\rm gc}}\rangle\;\bhat_{0}\bdot\ov{\nabla}\btimes\bhat_{0}\right)\ov{\cal F} \;+\; {\cal J}_{\rm gy} \left( \{ \delta{\cal S},\; \ov{\cal F}\}_{\rm gy} \;+\; \epsilon\;\frac{e}{c}\,\langle \delta \ov{A}_{1\|{\rm gc}}\rangle\; \pd{\ov{\cal F}}{\ov{p}_{\|}} \right) \;\equiv\; -\;\pd{}{\ov{\cal Z}^{a}}\left( {\cal F}_{\rm gy}\frac{}{}\delta\ov{\cal Z}^{a} \right),
  \label{eq:delta_F}
 \end{eqnarray}
 where we used
 \[ \delta{\cal J}_{\rm gy} \;=\; \epsilon\; \frac{e}{c}\,\langle\delta \ov{A}_{1\|{\rm gc}}\rangle\;\bhat_{0}\bdot\ov{\nabla}\btimes\bhat_{0} \;=\; \epsilon\; \frac{e}{c}\,\langle\delta \ov{A}_{1\|{\rm gc}}\rangle\;
 \pd{{\cal J}_{\rm gy}}{\ov{p}_{\|}}, \]
and the virtual displacement $\delta\ov{\cal Z}^{a}$ in extended gyrocenter phase space is defined as
 \begin{equation} 
 \delta\ov{\cal Z}^{a} \;\equiv\; \left\{ \ov{\cal Z}^{a},\frac{}{} \delta{\cal S} \right\}_{\rm gy} \;-\; \epsilon\;\frac{e}{c}\;\langle\delta \ov{A}_{1\|{\rm gc}}\rangle\;\bhat_{0}\bdot\left\{ \ov{\bf X},\frac{}{} \ov{\cal Z}^{a} 
 \right\}_{\rm gy}, 
 \label{eq:delta_Za} 
 \end{equation}
 and we used the relation $\partial_{a}(B_{\epsilon\|}^{*}\,\delta\ov{\cal Z}^{a}) = -\,\epsilon\,\langle\delta \ov{A}_{1\|{\rm gc}}\rangle\;\bhat_{0}\bdot\ov{\nabla}\btimes\bhat_{0} \equiv -\,\delta B_{\epsilon\|}^{*}$. Here,  
 $\delta{\cal S}$ generates the canonical part of a virtual transformation in extended phase space, and we used the gyrocenter Liouville property $B_{\epsilon\|}^{*}\,\{ \delta{\cal S},\; \ov{\cal F}\}_{\rm gy} = \partial_{a}(B_{\epsilon\|}^{*}\ov{\cal F}\,\{ \delta{\cal S}, \ov{\cal Z}^{a}\}_{\rm gy})$ to obtain the last expression in Eq.~\eqref{eq:delta_F}. We note that the term involving $\langle\delta \ov{A}_{1\|{\rm gc}}\rangle$ in Eq.~\eqref{eq:delta_F} appears as a result of the parallel-symplectic representation, i.e., because the first-order gyrocenter symplectic one-form \eqref{eq:Gamma_gy_1} is not zero (unlike in the Hamiltonian representation). A similar expression arises in the Eulerian variational formulation of guiding-center Vlasov-Maxwell theory \cite{Brizard_Tronci_2016}.

\subsection{Gyrokinetic variational principle}

By combining Eqs.~\eqref{eq:delta_Hgy} and \eqref{eq:Maxwell_var}-\eqref{eq:delta_F} into the gyrokinetic variational principle \eqref{eq:deltaA_gy}, we obtain
\begin{eqnarray}
\delta{\cal A}_{\rm gy} & = & -\;\sum \int d^{8}\ov{\cal Z} \left[ -\;{\cal H}_{\rm gy}\;\pd{}{\ov{\cal Z}^{a}}\left( {\cal F}_{\rm gy}\frac{}{}\delta\ov{\cal Z}^{a} \right) \;+\; {\cal F}_{\rm gy} \left( \epsilon\,e
\left\langle\delta\ov{\Phi}_{1{\rm gc}} \;-\; \frac{\Omega}{c}\,\pd{\ov{\vb{\rho}}_{0}}{\ov{\zeta}}\bdot\delta\ov{\bf A}_{1\bot{\rm gc}} \;-\; \epsilon\,\pounds_{1{\rm gy}}(\delta\ov{\psi}_{1{\rm gc}})\right\rangle \right) \right] \nonumber \\
 &  &-\; \int \frac{d^{4}x}{4\pi} \left( \epsilon^{2}\;\delta\Phi_{1}\;\nabla^{2}\Phi_{1} \;+\frac{}{} \epsilon\,\delta{\bf A}_{1}\bdot\nabla\btimes{\bf B} \right).
 \label{eq:delta_A_gy}
 \end{eqnarray}
 By integrating by parts the Vlasov variation term $\delta{\cal F}_{\rm gy}\,{\cal H}_{\rm gy}$ (here, we also assume that the components $\delta\ov{\cal Z}^{a}$ vanish on the phase-space integration boundary), we find
 \[ -\;\int d^{8}\ov{\cal Z}\; {\cal H}_{\rm gy}\;\pd{}{\ov{\cal Z}^{a}}\left( {\cal F}_{\rm gy}\frac{}{}\delta\ov{\cal Z}^{a} \right) \;=\; \int d^{8}\ov{\cal Z}\; {\cal J}_{\rm gy}\;\left[ \delta{\cal S}\;\left\{ \ov{\cal F},\frac{}{} 
 {\cal H}_{\rm gy}\right\}_{\rm gy} \;-\; \epsilon\,\frac{e}{c}\,\langle\delta \ov{A}_{1\|{\rm gc}}\rangle\; \ov{\cal F}\;\pd{H_{\rm gy}}{\ov{p}_{\|}} \right], \]
 where the space-time divergence terms contribute to the gyrokinetic Noether equation \eqref{eq:gy_Noether}. Hence, we rewrite Eq.~\eqref{eq:delta_A_gy} as
 \begin{eqnarray}
 \delta{\cal A}_{\rm gy} & = & -\; \int \frac{d^{4}x}{4\pi} \left( \epsilon^{2}\;\delta\Phi_{1}\;\nabla^{2}\Phi_{1} \;+\frac{}{}\epsilon\,\delta{\bf A}_{1}\bdot\nabla\btimes{\bf B} \right) \;-\; \sum \int d^{8}\ov{\cal Z} \;
 {\cal J}_{\rm gy}\;\delta{\cal S}\;\left\{ \ov{\cal F},\frac{}{} {\cal H}_{\rm gy}\right\}_{\rm gy} \nonumber \\
  &  &-\; \sum e\;\int d^{8}\ov{\cal Z}\;{\cal F}_{\rm gy}  \left[ \epsilon \left\langle\delta\ov{\Phi}_{1{\rm gc}} \;-\; \frac{\Omega}{c}\,\pd{\ov{\vb{\rho}}_{0}}{\ov{\zeta}}\bdot\delta\ov{\bf A}_{1\bot{\rm gc}} \right\rangle 
  \;-\; \epsilon\,\frac{e}{c}\,\langle\delta\ov{A}_{1\|{\rm gc}}\rangle\; \pd{H_{\rm gy}}{\ov{p}_{\|}} \;-\; \epsilon^{2}\left\langle\pounds_{1{\rm gy}}(\delta\ov{\psi}_{1{\rm gc}}) \right\rangle \right].
 \label{eq:delta_Agy_final}
 \end{eqnarray}
 In the next Section, we will derive the gyrokinetic Vlasov-Maxwell equations from the gyrokinetic variational principle $ \delta{\cal A}_{\rm gy} = 0$ for all variations $(\delta{\cal S}, \delta\Phi_{1}, \delta A_{1\|},
 \delta{\bf A}_{1\bot})$. 
 
 Lastly, we note that, once the gyrokinetic Vlasov-Maxwell equations from the gyrokinetic variational principle, the action variation \eqref{eq:delta_Agy_final} will be expressed in the form
 $\delta{\cal A}_{\rm gy} \equiv \int \delta{\cal L}_{\rm gy}\;d^{3}x\,dt$, from which we obtain the gyrokinetic Noether equation
 \begin{equation}
 \delta{\cal L}_{\rm gy} \;=\; \pd{}{t}\left( \sum\int {\cal F}_{\rm gy}\,\delta{\cal S}\;d^{4}\ov{P} \right) \;+\; \nabla\bdot\left[ \sum\int {\cal F}_{\rm gy}\,\delta{\cal S}\;\dot{\ov{\bf X}}_{\rm gy}\;d^{4}\ov{P} \;+\;
 \left(\epsilon^{2}\,\frac{\delta\Phi_{1}}{4\pi}\;\nabla\Phi_{1} - \epsilon\frac{\delta{\bf A}_{1}}{4\pi}\btimes{\bf B}\right) \right],
 \label{eq:gy_Noether}
 \end{equation}
 where $d^{4}\ov{P} = d\ov{p}_{\|} d\ov{J} d\ov{\zeta} d\ov{w}$ and the variations $(\delta{\cal S}, \delta\Phi_{1}, \delta{\bf A}_{1},  \delta{\cal L}_{\rm gy})$ will be expressed in terms of either a virtual time translation 
 $t \rightarrow t + \delta t$ or a virtual space translation ${\bf x} \rightarrow {\bf x} + \delta{\bf x}$, from which we will derive the gyrokinetic energy-momentum conservation laws, respectively. In the next Section, we will also derive the gyrokinetic energy conservation law (as well as the proof that energy is exactly conserved), while the gyrokinetic momentum conservation law will be derived elsewhere.
 
 \section{\label{sec:5}Gyrokinetic Vlasov-Maxwell Equations}

In this Section, the gyrokinetic Vlasov-Maxwell equations are derived from the gyrokinetic variational principle based on Eq.~\eqref{eq:delta_Agy_final}. Stationarity of the gyrokinetic action with respect to arbitrary variations $(\delta{\cal S},\delta\Phi_{1},\delta A_{1\|}, \delta{\bf A}_{1\bot})$ will yield, respectively, the gyrokinetic Vlasov equation, the gyrokinetic Poisson equation, the gyrokinetic parallel-Amp\`{e}re equation, and  
the gyrokinetic perpendicular-Amp\`{e}re equation, respectively. We also show that the gyrokinetic Vlasov-Maxwell equations possess an exact energy conservation law that is derived from the gyrokinetic
Noether equation \eqref{eq:gy_Noether}.

\subsection{Gyrokinetic Vlasov equation}

The extended gyrokinetic Vlasov equation $\{ \ov{\cal F}, {\cal H}_{\rm gy}\}_{\rm gy} = 0$ is obtained by requiring that $\delta{\cal A}_{\rm gy}$ vanishes for an arbitrary variation $\delta{\cal S}$. Upon integration with respect to $\ov{w}$, and using the gyrocenter Liouville property \eqref{eq:Liouville_gyPB}, we recover the divergence form of the gyrokinetic Vlasov equation:
 \begin{eqnarray}
 0 & = & \int d\ov{w}\;B_{\epsilon\|}^{*}\;\{ \ov{\cal F},\; {\cal H}_{\rm gy}\}_{\rm gy} \;=\; \int \pd{}{\ov{\cal Z}^{a}}\left( B_{\epsilon\|}^{*}\,\ov{\cal F}\;\left\{ \ov{\cal Z}^{a},\; {\cal H}_{\rm gy}\right\}_{\rm gy} \right) 
 \;d\ov{w} \nonumber \\
  & = & \pd{}{t} \left( \int  B_{\epsilon\|}^{*}\,\ov{\cal F}\; d\ov{w} \right) \;+\; \ov{\nabla}\bdot\left( \int  B_{\epsilon\|}^{*}\,\ov{\cal F}\;\dot{\ov{\bf X}}_{\rm gy} d\ov{w} \right) \;+\; \pd{}{\ov{p}_{\|}}\left(\int  B_{\epsilon\|}^{*}\,\ov{\cal F}\;\dot{\ov{p}}_{\|{\rm gy}} d\ov{w} \right) \nonumber \\
   & = & \pd{(B_{\epsilon\|}^{*}\,\ov{F})}{t}  \;+\; \ov{\nabla}\bdot\left( B_{\epsilon\|}^{*}\,\ov{F}\;\dot{\ov{\bf X}}_{\rm gy} \right) \;+\; \pd{}{\ov{p}_{\|}}\left(B_{\epsilon\|}^{*}\,\ov{F}\;\dot{\ov{p}}_{\|{\rm gy}} \right),
   \label{eq:gyVlasov_div}
  \end{eqnarray}
  where we used the definition \eqref{eq:calF_gy} of the extended gyrocenter Vlasov distribution. We obtain the gyrokinetic Vlasov equation 
\begin{equation}
\pd{\ov{F}}{t} \;+\; \dot{\ov{\bf X}}_{\rm gy}\;\bdot\ov{\nabla} \ov{F} \;+\; \dot{\ov{p}}_{\|{\rm gy}}\;\pd{\ov{F}}{\ov{p}_{\|}} \;=\; \left( \pd{\ov{F}}{t} \;-\; \epsilon\,\frac{e}{c}\,\pd{\langle\ov{A}_{1\|{\rm gc}}\rangle}{t}\;
\pd{\ov{F}}{\ov{p}_{\|}} \right) \;+\; \left\{ \ov{F},\frac{}{} H_{\rm gy} \right\}_{\rm gy} \;=\; 0
\label{eq:gy_Vlasov}
\end{equation}
from Eq.~\eqref{eq:gyVlasov_div} once we use the gyrocenter Liouville Theorem \eqref{eq:gy_Liouville}. The gyrokinetic Vlasov equation \eqref{eq:gy_Vlasov} describes the time evolution of the gyrocenter Vlasov distribution $\ov{F}(\ov{\bf X}, \ov{p}_{\|},t; \ov{J})$ in the $4+1$ reduced gyrocenter phase space $(\ov{\bf X}, \ov{p}_{\|}; \ov{J})$, where the gyrocenter Hamilton equations \eqref{eq:xdot_gy}-\eqref{eq:pdot_gy} are used as characteristics. Here, the term $\dot{\ov{\zeta}}_{\rm gy}\,\partial\ov{F}/\partial\ov{\zeta}$ is absent because $\partial\ov{F}/\partial\ov{\zeta} \equiv 0$ (i.e., the gyrocenter Vlasov distribution is gyroangle-independent at all orders in the guiding-center and gyrocenter orderings) and the gyrocenter gyroaction $\ov{J}$ appears as a gyrocenter invariant because 
$\dot{\ov{J}}_{\rm gy} \equiv 0$ (once again because the gyrocenter Hamiltonian is gyroangle-independent at all orders in the guiding-center and gyrocenter orderings). 

\subsection{Gyrokinetic Maxwell equations}

We now derive the gyrokinetic Maxwell equations from the gyrokinetic variational principle \eqref{eq:delta_Agy_final}. We begin with the gyrokinetic Poisson equation, which is derived from the stationarity of the gyrokinetic action functional \eqref{eq:delta_Agy_final} with respect to variations $\delta\Phi_{1}$:
\begin{equation}
\epsilon\,\nabla^{2}\Phi_{1}({\bf x},t) \;=\; -\,4\pi\,\sum e\;\int {\cal J}_{\rm gy}\;\ov{F}\;\left\langle \delta_{\rm gc}^{3} \;-\frac{}{} \epsilon\;\pounds_{1{\rm gy}}\left(\delta_{\rm gc}^{3}\right)\right\rangle\;d^{6}\ov{Z},
\label{eq:gy_Poisson_start}
\end{equation}
where we used the functional derivative \eqref{eq:fd_phi} and the first-order gyrocenter Lie-derivative of the (lowest-order) guiding-center charge distribution $e\,\delta^{3}_{\rm gc}$ is expressed as
\begin{eqnarray}
\pounds_{1{\rm gy}}\left(e\,\delta_{\rm gc}^{3}\right) & = & \{ S_{1},\; e\,\delta_{\rm gc}^{3} \}_{0} \;+\; \frac{e}{c}\,\left( \ov{\bf A}_{1\bot{\rm gc}}\bdot\left\{ \ov{\bf X} + \ov{\vb{\rho}}_{0},\frac{}{} 
e\,\delta_{\rm gc}^{3}  \right\}_{0} \;+\; \wt{A}_{1\|{\rm gc}}\;\bhat_{0}\bdot\left\{ \ov{\bf X},\frac{}{} e\,\delta_{\rm gc}^{3}  \right\}_{0} \right)  \nonumber \\
 & = & \{ S_{1},\; e\,\delta_{\rm gc}^{3} \}_{0} \;=\; e\,\left\{ S_{1},\frac{}{} \ov{\bf X} + \ov{\vb{\rho}}_{0}\right\}_{0}\bdot\ov{\nabla}\delta_{\rm gc}^{3} \;\equiv\; -\; \epsilon\;e\,\ov{\vb{\rho}}_{1{\rm gy}}\bdot\ov{\nabla} \delta_{\rm gc}^{3}.
 \label{eq:gy_push_Poisson}
\end{eqnarray}
By substituting Eq.~\eqref{eq:gy_push_Poisson} into the gyrokinetic Poisson equation \eqref{eq:gy_Poisson_start}, we obtain
\begin{equation}
\epsilon\,\nabla^{2}\Phi_{1} \;=\; -\,4\pi\,\sum \int {\cal J}_{\rm gy}\;\ov{F} \left( e\left\langle\delta_{\rm gc}^{3}\right\rangle \;+\frac{}{} \epsilon\;\left\langle e\,\ov{\vb{\rho}}_{1{\rm gy}}\bdot\ov{\nabla} \delta_{\rm gc}^{3}
\right\rangle\right) d^{6}\ov{Z} \;\equiv\; -\,4\pi \left( \varrho_{\rm gy} \;-\frac{}{} \nabla\bdot{\bf P}_{\rm gy}\right),
\label{eq:gy_Poisson}
\end{equation}
where $\varrho_{\rm gy}$ denotes the gyrocenter charge density and the gyrocenter polarization charge density $-\nabla\bdot{\bf P}_{\rm gy}$ is expressed in terms of the gyrocenter polarization ${\bf P}_{\rm gy}$, which includes a contribution from the first-order gyrocenter electric-dipole moment $e\,\ov{\vb{\rho}}_{1{\rm gy}}$ defined in Eq.~\eqref{eq:rho1_gy}. To lowest order in $\epsilon$, Eq.~\eqref{eq:gy_Poisson} yields the guiding-center quasineutrality condition $\varrho_{0{\rm gy}} = \nabla\bdot{\bf P}_{0{\rm gy}}$.

Next, we turn our attention to the gyrokinetic parallel-Amp\`{e}re equation, which is derived from the stationarity of the gyrokinetic action functional \eqref{eq:delta_Agy_final} with respect to variations 
$\delta A_{1\|}$:
\begin{equation}
\bhat_{0}\bdot\nabla\btimes{\bf B} \;=\; 4\pi\,\sum \frac{e}{c}\;\int {\cal J}_{\rm gy}\;\ov{F}\;\left[ \langle\delta_{\rm gc}^{3}\rangle\;\left(\frac{\ov{p}_{\|}}{m} \;+\; \epsilon^{2}\,\frac{e}{mc}\,\left\langle
\{ S_{1},\frac{}{} \wt{A}_{1\|{\rm gc}}\}_{0}\right\rangle \right) \;-\; \epsilon\;\left\langle\pounds_{1{\rm gy}} \left(\delta_{\rm gc}^{3}\;\frac{\ov{p}_{\|}}{m}\right) \right\rangle\right] d^{6}\ov{Z},
\label{eq:gy_Ampere_par}
\end{equation}
where we used the functional derivative \eqref{eq:fd_A_par} and the right side contains second-order gyrocenter ponderomotive corrections due to the parallel gyrocenter velocity \eqref{eq:H_p||}. We note that an important difference between the Hamiltonian and parallel-symplectic representations of gyrokinetic Vlasov-Maxwell theory involves the nonlinear gyrocenter parallel current associated with the ponderomotive contribution $\langle\{ S_{1}, \wt{A}_{1\|{\rm gc}}\}_{0}\rangle$, which is absent in the Hamiltonian representation.

On the other hand, the gyrokinetic perpendicular-Amp\`{e}re equation is derived from the stationarity of the gyrokinetic action functional \eqref{eq:delta_Agy_final} with respect to variations $\delta {\bf A}_{1\bot}$:
\begin{equation}
(\nabla\btimes{\bf B})_{\bot} \;=\; 4\pi\,\sum \frac{e}{c}\;\int {\cal J}_{\rm gy}\;\ov{F}\;\left[ \left\langle\delta_{\rm gc}^{3}\;\Omega\,\pd{\ov{\vb{\rho}}_{0}}{\ov{\zeta}}\right\rangle  \;-\; \epsilon\;\left\langle
\pounds_{1{\rm gy}} \left(\delta_{\rm gc}^{3}\;\Omega\,\pd{\ov{\vb{\rho}}_{0}}{\ov{\zeta}}\right) \right\rangle\right] d^{6}\ov{Z},
\label{eq:gy_Ampere_perp}
\end{equation}
where we used the functional derivative \eqref{eq:fd_A_perp} and we use the notation $(\nabla\btimes{\bf B})_{\bot} \equiv \bhat_{0}\btimes[(\nabla\btimes{\bf B})\btimes\bhat_{0}]$. We may combine these two equations to write the gyrokinetic Amp\`{e}re equation
\begin{eqnarray}
\nabla\btimes{\bf B}& = & 4\pi\,\sum \frac{e}{c}\;\int {\cal J}_{\rm gy}\;\ov{F}\;\left[ \left\langle \delta_{\rm gc}^{3}\,\left\{ {\bf X} + \ov{\vb{\rho}}_{0}, \ov{K}_{\rm gc}\right\}_{0} \right\rangle \;+\;
\epsilon^{2}\,\bhat_{0}\left( \frac{e}{mc}\,\left\langle\{ S_{1},\frac{}{} \wt{A}_{1\|{\rm gc}}\}_{0}\right\rangle \right) \right. \nonumber \\
 &  &\hspace*{1in}\left.-\; \epsilon\;\left\langle \pounds_{1{\rm gy}}\left(e\,\delta_{\rm gc}^{3}\frac{}{} \left\{ {\bf X} + \ov{\vb{\rho}}_{0}, \ov{K}_{\rm gc}\right\}_{0} \right)\right\rangle \right] d^{6}\ov{Z},
\label{eq:gy_Ampere}
\end{eqnarray}
where the first-order gyrocenter Lie-derivative of the guiding-center current distribution $e\,\delta_{\rm gc}^{3}\,\{{\bf X} + 
\ov{\vb{\rho}}_{0},\; \ov{K}_{\rm gc}\}_{0}$ is expressed as
\begin{eqnarray}
\pounds_{1{\rm gy}}\left(e\,\delta_{\rm gc}^{3}\frac{}{} \left\{ {\bf X} + \ov{\vb{\rho}}_{0}, \ov{K}_{\rm gc}\right\}_{0} \right) & = & -\,e\,\delta_{\rm gc}^{3} \left( \bhat_{0}\btimes\left\{ \ov{\bf X} + 
\ov{\vb{\rho}}_{0},\frac{}{} e\,\langle\ov{\psi}_{1{\rm gc}}\rangle \right\}_{0} \right)\btimes\bhat_{0} \nonumber \\
 &  &-\; \frac{d_{\rm gc}}{dt}\left(e\,\delta^{3}_{\rm gc}\frac{}{}\ov{\vb{\rho}}_{1{\rm gy}}\right) \;-\; \left(e\frac{}{}\ov{\vb{\rho}}_{1{\rm gy}}\btimes\left\{ \ov{\bf X} + \ov{\vb{\rho}}_{0}, \ov{K}_{\rm gc}\right\}_{0}\right)\btimes\ov{\nabla}\delta_{\rm gc}^{3}, 
\label{eq:gy_push_Ampere}
\end{eqnarray}
where we used the identity $\bhat_{0}\bdot\{ \ov{\bf X} + \ov{\vb{\rho}}_{0},e\,\langle\ov{\psi}_{1{\rm gc}}\rangle\}_{0} = -\;(e/mc)\langle \ov{A}_{1\|{\rm gc}}\rangle$. By substituting Eq.~\eqref{eq:gy_push_Ampere} into the gyrokinetic Amp\`{e}re equation \eqref{eq:gy_Ampere}, we obtain
\begin{eqnarray}
\nabla\btimes{\bf B} & = & \frac{4\pi}{c}\,\sum \int {\cal J}_{\rm gy}\;\ov{F} \left[ \left\langle e\,\delta_{\rm gc}^{3}\,\left\{ \ov{\bf X} + \ov{\vb{\rho}}_{0},\frac{}{} \ov{K}_{\rm gc}\right\}_{0} \right\rangle \;+\; 
\epsilon \left\langle e\,\delta_{\rm gc}^{3} \left( \bhat_{0}\btimes\left\{ \ov{\bf X} + \ov{\vb{\rho}}_{0},\frac{}{} e\,\langle\ov{\psi}_{1{\rm gc}}\rangle \right\}_{0} \right)\right\rangle\btimes\bhat_{0} \right. \nonumber \\
 &  &\left.+\; \epsilon^{2}\,\bhat_{0}\left( \frac{e}{mc}\,\left\langle\{ S_{1},\frac{}{} \wt{A}_{1\|{\rm gc}}\}_{0}\right\rangle \right) \;+\; \epsilon\;\frac{d_{\rm gc}}{dt} \left\langle e\,\delta^{3}_{\rm gc}\frac{}{}
 \ov{\vb{\rho}}_{1{\rm gy}}\right\rangle \;+\; \epsilon\left\langle\left(e\frac{}{}\ov{\vb{\rho}}_{1{\rm gy}}\btimes\left\{ \ov{\bf X} + \ov{\vb{\rho}}_{0},\frac{}{} \ov{K}_{\rm gc}\right\}_{0}\right)\btimes\ov{\nabla}
 \delta_{\rm gc}^{3}\right\rangle \right] d^{6}\ov{Z} \nonumber \\
  & \equiv & \frac{4\pi}{c} \left( {\bf J}_{\rm gy} \;+\; \pd{{\bf P}_{\rm gy}}{t} \;+\; c\,\nabla\btimes{\bf M}_{\rm gy} \right).
  \label{eq:Ampere_gy}
\end{eqnarray}
 Here, ${\bf J}_{\rm gy}$ denotes the gyrocenter current density, the gyrocenter polarization current density $\partial{\bf P}_{\rm gy}/\partial t$ is expressed in terms of the gyrocenter polarization ${\bf P}_{\rm gy}$, and the gyrocenter magnetization current density $c\,\nabla\btimes{\bf M}_{\rm gy}$ is expressed in terms of the gyrocenter magnetization ${\bf M}_{\rm gy}$. We note that the gyrocenter current density 
${\bf J}_{\rm gy}$ includes gyrocenter contributions up to second order in the perturbation parameter $\epsilon$, represented by the first three terms in the integrand on the right of Eq.~\eqref{eq:Ampere_gy}, respectively. To lowest order in $\epsilon$, Eq.~\eqref{eq:Ampere_gy} yields the guiding-center Amp\`{e}re equation $\nabla\btimes{\bf B}_{0} = 4\pi\,({\bf J}_{0{\rm gy}} + c\,\nabla\btimes{\bf M}_{0{\rm gy}})$.
    
\subsection{Energy conservation law}

We now show that the gyrokinetic Vlasov-Maxwell equations \eqref{eq:gy_Vlasov}, \eqref{eq:gy_Poisson} and \eqref{eq:gy_Ampere}, derived here in the parallel-symplectic representation, possess an exact energy conservation law. For this purpose, we return to the gyrokinetic Noether equation \eqref{eq:gy_Noether} and consider a virtual time translation $t \rightarrow t + \delta t$, with \cite{Brizard_2000a,Brizard_2000b}
\begin{equation}
\left. \begin{array}{rcl}
\delta{\cal S} & = & -\,\ov{w}\;\delta t \\
\delta\Phi_{1} & = & -\;\delta t\;\partial\Phi_{1}/\partial t \\
\delta{\bf A}_{1} & = & -\;\delta t\;\partial{\bf A}_{1}/\partial t
\end{array} \right\},
\end{equation}
and
\begin{eqnarray}
\delta{\cal L}_{\rm gy} & = & -\;\delta t\;\pd{}{t} \left[\frac{1}{8\pi}\;\left( \epsilon^{2}\;|\nabla\Phi_{1}|^{2} \;-\frac{}{} |{\bf B}_{0} + \epsilon\,\nabla\btimes{\bf A}_{1}|^{2} \right)\right] \nonumber \\
 & = & \delta t\;\pd{}{t} \left[ \frac{1}{8\pi}\;\left( \epsilon^{2}\;|\nabla\Phi_{1}|^{2} \;+\frac{}{} |{\bf B}_{0} + \epsilon\,\nabla\btimes{\bf A}_{1}|^{2} \right) \;+\; \epsilon\,\Phi_{1} \left( \frac{\epsilon}{4\pi}\;\nabla^{2}\Phi_{1}\right) \right] \;-\; \delta t\;\nabla\bdot\left[\pd{}{t}\left( \frac{\epsilon^{2}}{4\pi}\;\Phi_{1}\;\nabla\Phi_{1}\right)\right].
\end{eqnarray}
If we insert these expressions into the gyrokinetic Noether equation \eqref{eq:gy_Noether}, we obtain the local gyrokinetic energy conservation law
\begin{eqnarray}
0 & = & \pd{}{t}\left[ \sum\int {\cal F}_{\rm gy}\;\ov{w}\;d^{4}\ov{P} \;+\; \epsilon\,\Phi_{1} \left( \frac{\epsilon}{4\pi}\;\nabla^{2}\Phi_{1}\right) \;+\; \frac{1}{8\pi}\;\left( \epsilon^{2}\;|\nabla\Phi_{1}|^{2} \;+\frac{}{} 
|{\bf B}_{0} + \epsilon\,\nabla\btimes{\bf A}_{1}|^{2} \right) \right] \nonumber \\
 &  &+\; \nabla\bdot\left[ \sum\int {\cal F}_{\rm gy}\;\ov{w}\;\dot{\ov{\bf X}}_{\rm gy} \;d^{4}\ov{P} \;-\; \frac{\epsilon}{4\pi} \left( \epsilon\,\Phi_{1}\;\nabla\pd{\Phi_{1}}{t} \;+\; \pd{{\bf A}_{1}}{t}\bdot\nabla\btimes{\bf B}
 \right) \right],
 \end{eqnarray}
 whose form is identical to the case of the Hamiltonian representation \cite{Brizard_2000b}. When this local conservation is integrated over the entire spatial volume (assuming that the fields $({\cal F}_{\rm gy},
 \Phi_{1}, {\bf A}_{1})$ vanish on the integration boundary, we obtain the global gyrokinetic energy conservation law $d{\cal E}_{\rm gy}/dt = 0$, where the total gyrokinetic energy is
 \begin{eqnarray}
 {\cal E}_{\rm gy} & = &  \sum\int {\cal F}_{\rm gy}\;\ov{w}\;d^{6}\ov{Z}\,d\ov{w} \;+\; \int d^{3}x \left[ \epsilon\,\Phi_{1} \left( \frac{\epsilon}{4\pi}\;\nabla^{2}\Phi_{1}\right) \;+\; \frac{1}{8\pi}\;\left( 
 \epsilon^{2}\;|\nabla\Phi_{1}|^{2} \;+\frac{}{} |{\bf B}_{0} + \epsilon\,\nabla\btimes{\bf A}_{1}|^{2} \right) \right] \nonumber \\
 & = & \sum \int {\cal J}_{\rm gy}\;\ov{F} \left[ H_{\rm gy} \;-\; e\,\epsilon \left( \langle\Phi_{1{\rm gc}}\rangle \;-\; \epsilon\;\left\langle \{ S_{1},\frac{}{} \Phi_{1{\rm gc}}\}_{0}\right\rangle \right) \right]\; d^{6}\ov{Z} \nonumber \\
  &  &+\; \int \frac{d^{3}x}{8\pi} \left( \epsilon^{2}\;|\nabla\Phi_{1}|^{2} \;+\frac{}{} |{\bf B}_{0} + \epsilon\,\nabla\btimes{\bf A}_{1}|^{2} \right),
  \label{eq:E_gy}
  \end{eqnarray}
  where we inserted the gyrokinetic Poisson equation \eqref{eq:gy_Poisson}. We now evaluate the time derivative of the total gyrokinetic energy \eqref{eq:E_gy}:
  \begin{eqnarray*}
 \frac{d{\cal E}_{\rm gy}}{dt} & = & \sum \int \pd{({\cal J}_{\rm gy}\;\ov{F})}{t} \left[ H_{\rm gy} \;-\; e\,\epsilon \left( \langle\Phi_{1{\rm gc}}\rangle \;-\; \epsilon\;\left\langle \{ S_{1},\frac{}{} \Phi_{1{\rm gc}}\}_{0}\right\rangle \right) \right]\; d^{6}\ov{Z} \\
 &  &+\; \sum \int {\cal J}_{\rm gy}\;\ov{F} \left[ \pd{H_{\rm gy}}{t} \;-\; e\,\epsilon \left( \pd{\langle\Phi_{1{\rm gc}}\rangle}{t} \;-\; \epsilon\;\left\langle \left\{ \pd{S_{1}}{t},\Phi_{1{\rm gc}}\right\}_{0} + \left\{ S_{1},
 \pd{\Phi_{1{\rm gc}}}{t} \right\}_{0}\right\rangle \right) 
 \right]\; d^{6}\ov{Z} \\
  &  &+\; \int d^{3}x \left[ \epsilon\;\Phi_{1}\left(-\,\frac{\epsilon}{4\pi}\;\nabla^{2}\pd{\Phi_{1}}{t}\right) \;+\; \frac{\epsilon}{4\pi}\,\pd{{\bf A}_{1}}{t}\bdot\nabla\btimes{\bf B} \right],
  \end{eqnarray*} 
  which becomes 
  \begin{equation}
  \frac{d{\cal E}_{\rm gy}}{dt} \;=\; \sum \int \left[ \pd{({\cal J}_{\rm gy}\;\ov{F})}{t}\;H_{\rm gy} \;+\; {\cal J}_{\rm gy}\;\ov{F} \left( \epsilon\,\frac{e}{c}\pd{\langle A_{1\|{\rm gc}}\rangle}{t}\;\pd{H_{\rm gy}}{\ov{p}_{\|}}\right)
  \right] d^{6}\ov{Z},
  \label{eq:Egy_dot}
  \end{equation}
  after inserting the gyrokinetic Maxwell equations \eqref{eq:gy_Poisson} and \eqref{eq:gy_Ampere}. By integrating the second term by parts, we find
\[ \int  {\cal J}_{\rm gy}\;\ov{F} \left( \epsilon\,\frac{e}{c}\pd{\langle A_{1\|{\rm gc}}\rangle}{t}\;\pd{H_{\rm gy}}{\ov{p}_{\|}}\right) d^{6}\ov{Z}  \;=\; -\; \int \left[ \ov{F}\,H_{\rm gy}\;\left( \epsilon\,\frac{e}{c}
\pd{\langle A_{1\|{\rm gc}}\rangle}{t}\;\pd{{\cal J}_{\rm gy}}{\ov{p}_{\|}}\right) \;+\; {\cal J}_{\rm gy}\, H_{\rm gy} \left( \epsilon\,\frac{e}{c}\pd{\langle A_{1\|{\rm gc}}\rangle}{t}\;\pd{\ov{F}}{\ov{p}_{\|}}\right)\right], \]
so that Eq.~\eqref{eq:Egy_dot} becomes
 \begin{equation}
  \frac{d{\cal E}_{\rm gy}}{dt} \;=\; \sum \int \left[ {\cal J}_{\rm gy}\, H_{\rm gy}  \left( \pd{\ov{F}}{t} \;-\; \epsilon\,\frac{e}{c}\pd{\langle A_{1\|{\rm gc}}\rangle}{t}\;\pd{\ov{F}}{\ov{p}_{\|}}\right) \;+\;
  \ov{F}\, H_{\rm gy}  \left( \pd{{\cal J}_{\rm gy}}{t} \;-\; \epsilon\,\frac{e}{c}\pd{\langle A_{1\|{\rm gc}}\rangle}{t}\;\pd{{\cal J}_{\rm gy}}{\ov{p}_{\|}}\right) \right] d^{6}\ov{Z}.  
 \end{equation}
 Lastly, using the gyrokinetic Vlasov equation \eqref{eq:gy_Vlasov}, the first term vanishes 
 \[ \int {\cal J}_{\rm gy}\, H_{\rm gy}  \left( \pd{\ov{F}}{t} \;-\; \epsilon\,\frac{e}{c}\pd{\langle A_{1\|{\rm gc}}\rangle}{t}\;\pd{\ov{F}}{\ov{p}_{\|}}\right) d^{6}\ov{Z} \;=\; -\;\int {\cal J}_{\rm gy}\, H_{\rm gy}\;
 \left\{ \ov{F},\frac{}{} H_{\rm gy} \right\}_{\rm gy} \;=\; 0, \]
since ${\cal J}_{\rm gy}\, H_{\rm gy}\;\{ \ov{F}, H_{\rm gy}\}_{\rm gy} = {\cal J}_{\rm gy}\,\{ \ov{F}\,H_{\rm gy}, H_{\rm gy}\}_{\rm gy}$ can be written as an exact phase-space divergence. Hence, the exact gyrokinetic energy conservation law $d{\cal E}_{\rm gy}/dt = 0$ is satisfied by the gyrokinetic Vlasov-Maxwell equations \eqref{eq:gy_Vlasov}, \eqref{eq:gy_Poisson} and \eqref{eq:gy_Ampere}, provided the gyrocenter Liouville condition \eqref{eq:Jac_gy_dot} is satisfied.

 \section{\label{sec:6}Summary}
 
 In this paper, we have shown how the nonlinear gyrokinetic Vlasov-Maxwell equations \eqref{eq:gy_Vlasov}, \eqref{eq:gy_Poisson}, and \eqref{eq:gy_Ampere} in the parallel-symplectic representation can be self-consistently derived from an Eulerian variational principle based on the gyrokinetic action functional \eqref{eq:delta_A_gy}. From the gyrokinetic Noether equation \eqref{eq:gy_Noether} obtained from the variational principle, we derived an exact energy conservation law for the gyrokinetic Vlasov-Maxwell equations \eqref{eq:gy_Vlasov}, \eqref{eq:gy_Poisson}, and \eqref{eq:gy_Ampere} in the parallel-symplectic representation. 
 
 We now summarize the differences between the Hamiltonian and parallel-symplectic representations as follows. First, in the parallel-symplectic representation, both the gyrocenter Poisson bracket 
 \eqref{eq:PB_gy} and the gyrocenter Jacobian \eqref{eq:Jac_gy} include contributions from the gyroangle-averaged parallel component $\langle \ov{A}_{1\|{\rm gc}}\rangle$. Despite these time-dependent contributions, the gyrocenter Poisson bracket and the gyrocenter Jacobian still satisfy the gyrocenter Liouville Theorem \eqref{eq:Liouville_gyPB} and property \eqref{eq:gy_Liouville}, which play a crucial role in the variational derivation of the gyrokinetic Vlasov-Maxwell equations \eqref{eq:gy_Vlasov}, \eqref{eq:gy_Poisson}, and \eqref{eq:gy_Ampere}. Second, in the parallel-symplectic representation, the gyrocenter parallel-Amp\`{e}re equation \eqref{eq:gy_Ampere_par} contains a second-order contribution that is derived from the second-order gyrocenter ponderomotive Hamiltonian [see Eq.~\eqref{eq:H_p||}]. In the Hamiltonian representation, this second-order ponderomotive contribution does not appear in the the gyrocenter parallel-Amp\`{e}re equation because the constrained variation \eqref{eq:delta_F} is now simply expressed as 
 $\delta\ov{\cal F} \equiv \{\delta{\cal S},\;\ov{\cal F}\}_{0}$.
  
Lastly, we note that the truncated ($\delta f$) gyrokinetic Vlasov-Maxwell equations can be derived in the parallel-symplectic representation from the truncated gyrokinetic variational principle \cite{Brizard_2010} 
\begin{eqnarray}
{\cal A}_{\rm trgy}[\ov{F}_{1}, \Phi_{1}, {\bf A}_{1}] & = & -\;\sum  \int \left[{\cal F}_{\rm trgy}\,{\cal H}_{\rm trgy} \;-\; \frac{\epsilon^{2}}{2}\;{\cal J}_{0{\rm gy}}\,\ov{F}_{0}\;\left\langle \pounds_{1{\rm gy}}\left(
e\frac{}{} \ov{\psi}_{1{\rm gc}}\right)\right\rangle \right]d^{8}\ov{\cal Z} \nonumber \\
 &  &+\; \int \frac{d^{3}x\,dt}{8\pi} \left( \epsilon^{2}\;|\nabla\Phi_{1}|^{2} \;-\frac{}{} |{\bf B}_{0} + \epsilon\,\nabla\btimes{\bf A}_{1}|^{2} \right),
\label{eq:A_trgy}
\end{eqnarray}
where the truncated extended gyrocenter Vlasov density ${\cal F}_{\rm trgy} = {\cal J}_{\rm gy}\,(\ov{F}_{0} + \epsilon\,\ov{F}_{1})\;\delta(\ov{w} - H_{\rm trgy})$ is defined from Eq.~\eqref{eq:calF_gy} by truncating the gyrocenter Hamiltonian \eqref{eq:H_gy} at first order in $\epsilon$: $H_{\rm trgy} \equiv H_{0{\rm gy}} + \epsilon\,H_{1{\rm gy}}$ and expanding the gyrocenter Vlasov distribution function $\ov{F} = \ov{F}_{0} + \epsilon\,\ov{F}_{1}$ in terms of a time-independent reference gyrocenter Vlasov distribution $\ov{F}_{0}$ and a time-dependent departure $\ov{F}_{1}$ from $\ov{F}_{0}$ that is driven by linear and nonlinear effects. Here, the zeroth-order gyrocenter Jacobian is ${\cal J}_{0{\rm gy}} = (e/c) B_{0\|}^{*}$ and $\ov{F}_{0}$ is assumed to satisfy the zeroth-order gyrocenter Vlasov equation $\{ \ov{F}_{0},\;\ov{H}_{0{\rm gy}}\}_{0} = 0$. Because the second-order gyrocenter ponderomotive Hamiltonian is now associated with the gyrocenter Vlasov distribution $\ov{F}_{0}$, however, it does not contribute a second-order contribution to the gyrocenter parallel-Amp\`{e}re equation (see Ref.~\cite{Tronko_2016} for additional details).

\acknowledgments

This paper is offered in celebration of We-li Lee's pioneering contributions in the development and applications of gyrokinetic particle-in-cell simulation techniques of magnetized plasmas. Nearly thirty years ago, as a graduate student, I had the honor of working with Wei-li Lee and T.S. Hahm on Ref.~\cite{HLB_1988}. Since then, I have greatly appreciated Wei-li's insightful questions about my work on gyrokinetic theory, as well as his unwavering interest and support. I also wish to acknowledge the work of a tenacious referee who made me discover new aspects of gyrokinetic theory (e.g., second-order effects in the gyrocenter parallel-Amp\`{e}re equation). The present work was partially funded by a grant from the U.~S.~Dept.~of Energy under contract DE-SC0014032.

\appendix

\section{\label{sec:App_A}Gyrocenter Liouville Property}

In this Appendix, we prove the gyrocenter Liouville property \eqref{eq:Liouville_gyPB} of the gyrocenter Poisson bracket \eqref{eq:PB_gy}. First, we write the expression
\begin{eqnarray}
\frac{1}{B_{\epsilon\|}^{*}}\pd{}{\ov{\cal Z}^{a}}\left(B_{\epsilon\|}^{*}\,\ov{\cal F}\frac{}{}\{ \ov{\cal Z}^{a},\; \ov{\cal G}\}_{\rm gy}\right) & = & \frac{1}{B_{\epsilon\|}^{*}}\pd{}{\ov{\cal Z}^{a}}\left(B_{\epsilon\|}^{*}\,
\ov{\cal F}J_{\rm gy}^{ab}\;\pd{\ov{\cal G}}{\ov{\cal Z}^{b}} \right) \nonumber \\
 & = & \frac{1}{B_{\epsilon\|}^{*}}\pd{}{\ov{\cal Z}^{a}}\left(B_{\epsilon\|}^{*}\frac{}{}J_{\rm gy}^{ab}\right)\;\ov{\cal F}\,\pd{\ov{\cal G}}{\ov{\cal Z}^{b}} \;+\; \{ \ov{\cal F},\; \ov{\cal G}\}_{\rm gy} \;+\;
 \ov{\cal F}\;\left( J_{\rm gy}^{ab}\;\frac{\partial^{2}\ov{\cal G}}{\partial\ov{\cal Z}^{a}\partial\ov{\cal Z}^{b}}\right) \nonumber \\
  & = & \frac{1}{B_{\epsilon\|}^{*}}\pd{}{\ov{\cal Z}^{a}}\left(B_{\epsilon\|}^{*}\frac{}{}J_{\rm gy}^{ab}\right)\;\ov{\cal F}\,\pd{\ov{\cal G}}{\ov{\cal Z}^{b}} \;+\; \{ \ov{\cal F},\; \ov{\cal G}\}_{\rm gy},
  \label{eq:A1}
\end{eqnarray}
where the last line is obtained by using the antisymmetry of the gyrocenter Poisson tensor $J_{\rm gy}^{ab} = \{ \ov{\cal Z}^{a}, \ov{\cal Z}^{b}\}_{\rm gy}$ and the symmetry of the second derivative $\partial^{2}_{ab}\ov{\cal G}$, which yields $J_{\rm gy}^{ab}\,\partial^{2}_{ab}\ov{\cal G} \equiv 0$. Equation \eqref{eq:A1}, therefore, yields the gyrocenter Liouville property 
\begin{equation}
\frac{1}{B_{\epsilon\|}^{*}}\pd{}{\ov{\cal Z}^{a}}\left(B_{\epsilon\|}^{*}\,\ov{\cal F}\frac{}{}\{ \ov{\cal Z}^{a},\; \ov{\cal G}\}_{\rm gy}\right)  \;=\; \{ \ov{\cal F},\; \ov{\cal G}\}_{\rm gy} \;=\; \frac{1}{B_{\epsilon\|}^{*}}
\pd{}{\ov{\cal Z}^{b}}\left(B_{\epsilon\|}^{*}\,\frac{}{}\{ \ov{\cal F},\; \ov{\cal Z}^{b}\}_{\rm gy}\;\ov{\cal G}\right) 
\label{eq:A2}
\end{equation}
provided the gyrocenter Liouville identities
\begin{equation}
0 = \pd{}{\ov{\cal Z}^{a}}\left(B_{\epsilon\|}^{*}\frac{}{}J_{\rm gy}^{ab}\right) = \ov{\nabla}\bdot\left( B_{\epsilon\|}^{*}\frac{}{}\{ \ov{\bf X},\; \ov{\cal Z}^{b}\}_{\rm gy}\right) + \pd{}{\ov{p}_{\|}}\left( B_{\epsilon\|}^{*}\frac{}{}\{ \ov{p}_{\|},\; \ov{\cal Z}^{b}\}_{\rm gy}\right) + \pd{}{\ov{J}}\left( B_{\epsilon\|}^{*}\frac{}{}\{ \ov{J},\; \ov{\cal Z}^{b}\}_{\rm gy}\right) + \pd{}{t}\left( B_{\epsilon\|}^{*}\frac{}{}\{ t,\; \ov{\cal Z}^{b}\}_{\rm gy}\right)
\label{eq:A3}
\end{equation}
are satisfied for all $\ov{\cal Z}^{b}$, where we have used the fact that $\partial_{\ov{\zeta}}(B_{\epsilon\|}^{*}\,\{ \ov{\zeta},\; \ov{\cal Z}^{b}\}_{\rm gy}) = 0 = \partial_{\ov{w}}(B_{\epsilon\|}^{*}\,\{ \ov{w},\; 
\ov{\cal Z}^{b}\}_{\rm gy})$. 

To prove the gyrocenter Liouville identities \eqref{eq:A3}, we write the components
\begin{eqnarray*}
B_{\epsilon\|}^{*}\frac{}{}\{ \ov{\bf X},\; \ov{\cal Z}^{b}\}_{\rm gy} & = & {\bf B}_{\epsilon}^{*}\;\pd{\ov{\cal Z}^{b}}{\ov{p}_{\|}} \;+\; \frac{c\bhat_{0}}{e}\btimes\ov{\nabla}_{0}^{*}\ov{\cal Z}^{b}, \\
B_{\epsilon\|}^{*}\frac{}{}\{ \ov{p}_{\|},\; \ov{\cal Z}^{b}\}_{\rm gy} & = & -\;{\bf B}_{\epsilon}^{*}\bdot\left[ \ov{\nabla}_{0}^{*}\ov{\cal Z}^{b} \;-\; \epsilon\,\frac{e\bhat_{0}}{c} \left( 
\pd{\langle\ov{A}_{1\|{\rm gc}}\rangle}{\ov{J}}\,\pd{\ov{\cal Z}^{b}}{\ov{\zeta}} + \pd{\langle\ov{A}_{1\|{\rm gc}}\rangle}{t}\,\pd{\ov{\cal Z}^{b}}{\ov{w}} \right) \right],
\end{eqnarray*}
with $\{ \ov{J},\; \ov{\cal Z}^{b}\}_{\rm gy} = -\;\partial\ov{\cal Z}^{b}/\partial\ov{\zeta}$ and $\{ t,\; \ov{\cal Z}^{b}\}_{\rm gy} = -\;\partial\ov{\cal Z}^{b}/\partial\ov{w}$, where we note that the derivatives 
$\partial_{a}\ov{\cal Z}^{b} \equiv \delta_{a}^{b}$ are constants. Next, we write the derivatives
\begin{eqnarray*}
\ov{\nabla}\bdot\left(B_{\epsilon\|}^{*}\frac{}{}\{ \ov{\bf X},\; \ov{\cal Z}^{b}\}_{\rm gy}\right) & = & \ov{\nabla}\btimes\left(\frac{c\bhat_{0}}{e}\right)\bdot\ov{\nabla}_{0}^{*}\ov{\cal Z}^{b} \;-\; \frac{c\bhat_{0}}{e}
\bdot\ov{\nabla}\btimes{\bf R}_{0}^{*}\;\pd{\ov{\cal Z}^{b}}{\ov{\zeta}}, \\
\pd{}{\ov{p}_{\|}}\left(B_{\epsilon\|}^{*}\frac{}{}\{ \ov{p}_{\|},\; \ov{\cal Z}^{b}\}_{\rm gy} \right) & = & -\;\ov{\nabla}\btimes\left(\frac{c\bhat_{0}}{e}\right)\bdot\left[ \ov{\nabla}_{0}^{*}\ov{\cal Z}^{b} \;-\; \epsilon\,\frac{e\bhat_{0}}{c} \left( \pd{\langle\ov{A}_{1\|{\rm gc}}\rangle}{\ov{J}}\,\pd{\ov{\cal Z}^{b}}{\ov{\zeta}} + \pd{\langle\ov{A}_{1\|{\rm gc}}\rangle}{t}\,\pd{\ov{\cal Z}^{b}}{\ov{w}} \right) \right], \\
\pd{}{\ov{J}}\left(B_{\epsilon\|}^{*}\frac{}{}\{ \ov{J},\; \ov{\cal Z}^{b}\}_{\rm gy} \right) & = & -\;\ov{\nabla}\btimes\left(-\,{\bf R}_{0}^{*} \;+\; \epsilon\,\frac{e\bhat_{0}}{c}\;\pd{\langle\ov{A}_{1\|{\rm gc}}\rangle}{\ov{J}}\right)\bdot\frac{c\bhat_{0}}{e}\;\pd{\ov{\cal Z}^{b}}{\ov{\zeta}}, \\
\pd{}{t}\left(B_{\epsilon\|}^{*}\frac{}{}\{ t,\; \ov{\cal Z}^{b}\}_{\rm gy} \right) & = & -\;\ov{\nabla}\btimes\left(\epsilon\,\bhat_{0}\;\pd{\langle\ov{A}_{1\|{\rm gc}}\rangle}{t}\right)\bdot\bhat_{0}\;\;\pd{\ov{\cal Z}^{b}}{\ov{w}}.
\end{eqnarray*}
which yield exact cancellations when inserted in Eq.~\eqref{eq:A3}. Hence, since the gyrocenter Liouville identities \eqref{eq:A3} are satisfied, then we have proved the gyrocenter Liouville property \eqref{eq:A2}.

\section{\label{sec:App_B}Gyrocenter Ponderomotive Hamiltonian}

In this Appendix, we derive the gyrocenter Hamiltonian identity \eqref{eq:2gy_id} as follows. First, using $\ov{\bf E}_{1{\rm gc}} = -\,\ov{\nabla}\ov{\Phi}_{1{\rm gc}} - c^{-1}\partial_{t}\ov{\bf A}_{1{\rm gc}}$ and 
$\ov{\bf B}_{1{\rm gc}} = \ov{\nabla}\btimes\ov{\bf A}_{1{\rm gc}}$, we write
\begin{eqnarray}
e\,\left\{ S_{1},\frac{}{} \ov{\bf X} + \ov{\vb{\rho}}_{0}\right\}_{0}\bdot\ov{\bf E}_{1{\rm gc}} & = & -\;e\,\{ S_{1},\; \ov{\bf X} + \ov{\vb{\rho}}_{0}\}_{0}\bdot\ov{\nabla}\ov{\Phi}_{1{\rm gc}} \;-\; \frac{e}{c}\left\{ S_{1},\frac{}{} 
\ov{\bf X} + \ov{\vb{\rho}}_{0}\right\}_{0}\bdot\pd{\ov{\bf A}_{1{\rm gc}}}{t} \nonumber \\
 & = & -\;\{ S_{1},\; e\ov{\Phi}_{1{\rm gc}}\}_{0} \;+\; \frac{e}{c}\pd{}{t}\left( \ov{\bf A}_{1{\rm gc}}\bdot\{ \ov{\bf X} + \ov{\vb{\rho}}_{0},\frac{}{} S_{1}\}_{0}\right) \;-\; \frac{e}{c}\ov{\bf A}_{1{\rm gc}}\bdot\left\{ \ov{\bf X} + 
 \ov{\vb{\rho}}_{0},\; \pd{S_{1}}{t}\right\}_{0}
 \label{eq:E_gc}
 \end{eqnarray}
 and
 \begin{eqnarray}
e\,\left\{ S_{1},\frac{}{} \ov{\bf X} + \ov{\vb{\rho}}_{0}\right\}_{0}\bdot \left(\frac{1}{c}\{ \ov{\bf X} + \ov{\vb{\rho}}_{0},\frac{}{} \ov{K}_{\rm gc}\}_{0}\btimes \ov{\bf B}_{1{\rm gc}}\right) & = & \frac{e}{c}\,\{ S_{1},\; \ov{\bf X} + \ov{\vb{\rho}}_{0}\}_{0}\bdot\ov{\nabla}\ov{\bf A}_{1{\rm gc}}\bdot\{ \ov{\bf X} + \ov{\vb{\rho}}_{0},\; \ov{K}_{\rm gc}\}_{0} \nonumber \\
 &  &-\; \frac{e}{c}\,\{ \ov{\bf X} + \ov{\vb{\rho}}_{0},\; \ov{K}_{\rm gc}\}_{0} \bdot\ov{\nabla}\ov{\bf A}_{1{\rm gc}}\bdot\{ S_{1},\; \ov{\bf X} + \ov{\vb{\rho}}_{0}\}_{0} \nonumber \\
  & = & \left\{ S_{1},\;  \frac{e}{c}\,\ov{\bf A}_{1{\rm gc}}\bdot\{ \ov{\bf X} + \ov{\vb{\rho}}_{0},\; \ov{K}_{\rm gc}\}_{0}\right\}_{0} \nonumber \\
   &  &-\; \frac{e}{c}\ov{\bf A}_{1{\rm gc}}\bdot\left\{ S_{1},\frac{}{} \{\ov{\bf X} + \ov{\vb{\rho}}_{0},\; \ov{K}_{\rm gc}\}_{0}\right\}_{0} \nonumber \\
   &  &+\; \frac{e}{c}\{ \ov{K}_{\rm gc},\;\ov{\bf A}_{1{\rm gc}}\}_{0} \bdot\{ S_{1},\; \ov{\bf X} + \ov{\vb{\rho}}_{0}\}_{0},
   \label{eq:B_gc}
 \end{eqnarray}
 where we used the identity $\{ F,\; \ov{\bf X} + \ov{\vb{\rho}}_{0}\}_{0}\bdot\ov{\nabla}(\cdots)_{1{\rm gc}} \equiv \{ F,\; (\cdots)_{1{\rm gc}}\}_{0}$. Next, we use the Jacobi identity \eqref{eq:Jac_gy} for the Poisson bracket  $\{ \;,\; \}_{0}$ to obtain
 \begin{eqnarray}
-\; \frac{e}{c}\ov{\bf A}_{1{\rm gc}}\bdot\left\{ S_{1},\frac{}{} \{\ov{\bf X} + \ov{\vb{\rho}}_{0},\; \ov{K}_{\rm gc}\}_{0}\right\}_{0} & = & -\; \frac{e}{c}\ov{\bf A}_{1{\rm gc}}\bdot\left( \left\{ \ov{\bf X} + \ov{\vb{\rho}}_{0},\frac{}{} 
\{S_{1},\; \ov{K}_{\rm gc}\}_{0}\right\}_{0} + \left\{ \{ S_{1},\; \ov{\bf X} + \ov{\vb{\rho}}_{0}\}_{0},\frac{}{} \ov{K}_{\rm gc}\right\}_{0}  \right) \\
 & = & -\; \frac{e}{c}\ov{\bf A}_{1{\rm gc}}\bdot\left\{ \ov{\bf X} + \ov{\vb{\rho}}_{0},\frac{}{} \{S_{1},\; \ov{K}_{\rm gc}\}_{0}\right\}_{0} \;+\; \frac{e}{c} \left\{ \ov{\bf A}_{1{\rm gc}}\bdot\{ \ov{\bf X} + \ov{\vb{\rho}}_{0},\; 
 S_{1}\}_{0},\frac{}{} \ov{K}_{\rm gc} \right\}_{0} \nonumber \\
  & &-\; \frac{e}{c}\{ \ov{K}_{\rm gc},\;\ov{\bf A}_{1{\rm gc}}\}_{0} \bdot\{ S_{1},\; \ov{\bf X} + \ov{\vb{\rho}}_{0}\}_{0}, \nonumber
\end{eqnarray}
so that by adding Eqs.~\eqref{eq:E_gc}-\eqref{eq:B_gc}, we obtain
\begin{eqnarray}
e\,\left\{ S_{1},\frac{}{} \ov{\bf X} + \ov{\vb{\rho}}_{0}\right\}_{0}\bdot\left(\ov{\bf E}_{1{\rm gc}} + \frac{1}{c}\{ \ov{\bf X} + \ov{\vb{\rho}}_{0},\frac{}{} \ov{K}_{\rm gc}\}_{0}\btimes \ov{\bf B}_{1{\rm gc}}\right) & = & -\; 
\{ S_{1},\; e\,\ov{\psi}_{1{\rm gc}}\}_{0} \;-\; \frac{e}{c}\ov{\bf A}_{1{\rm gc}}\bdot\left\{ \ov{\bf X} + \ov{\vb{\rho}}_{0},\frac{}{} \{S_{1},\; {\cal H}_{0{\rm gc}}\}_{0}\right\}_{0} \nonumber \\
 &  &+\; \frac{e}{c}\;\frac{d_{\rm gc}}{dt}\left( \ov{\bf A}_{1{\rm gc}}\bdot\{ \ov{\bf X} + \ov{\vb{\rho}}_{0},\frac{}{} S_{1}\}_{0} \right).
\end{eqnarray}
Lastly, by using Eq.~\eqref{eq:S1_dot}, we finally obtain the gyroangle-averaged expression \eqref{eq:2gy_id}:
\begin{eqnarray}
\left\langle\left\{ S_{1},\frac{}{} \ov{\bf X} + \ov{\vb{\rho}}_{0}\right\}_{0}\bdot\left(e\,\ov{\bf E}_{1{\rm gc}} + \frac{e}{c}\{ \ov{\bf X} + \ov{\vb{\rho}}_{0},\frac{}{} \ov{K}_{\rm gc}\}_{0}\btimes \ov{\bf B}_{1{\rm gc}}\right) \right\rangle & = & -\; \left\langle\left\{ S_{1},\frac{}{} \{ S_{1},\; \ov{\cal H}_{0{\rm gc}}\}_{0} \right\}_{0} \right\rangle \nonumber \\
 &  &-\; \frac{e}{c}\left\langle\ov{\bf A}_{1{\rm gc}}\bdot\left\{ \ov{\bf X} + \ov{\vb{\rho}}_{0},\frac{}{} e\,(\ov{\psi}_{1{\rm gc}} - \langle\ov{\psi}_{1{\rm gc}}\rangle)\right\}_{0} \right\rangle \nonumber \\
 &  &+\; \frac{e}{c}\;\frac{d_{\rm gc}}{dt}\left\langle \ov{\bf A}_{1{\rm gc}}\bdot\{ \ov{\bf X} + \ov{\vb{\rho}}_{0},\frac{}{} S_{1}\}_{0} \right\rangle \nonumber \\
  & \equiv & \frac{e}{c}\,\left\langle \ov{\bf A}_{1{\rm gc}}\bdot\left\{ \ov{\bf X} + \ov{\vb{\rho}}_{0},\frac{}{} e\,\langle\ov{\psi}_{1{\rm gc}}\rangle\right\}_{0} \right\rangle + \frac{e^{2}}{mc^{2}}\;
 \left\langle|\ov{\bf A}_{1{\rm gc}}|^{2}\right\rangle \nonumber \\
 &  &-\; \left\langle \left\{ S_{1},\frac{}{} \{ S_{1},\; \ov{\cal H}_{0{\rm gc}}\}_{0}\right\}_{0}\right\rangle \;+\; \frac{e}{c}\,\frac{d_{\rm gc}}{dt}\langle\chi_{2}\rangle,
\end{eqnarray}
where we used the guiding-center identity $\{ \ov{\bf X} + \ov{\vb{\rho}}_{0},\; \{ \ov{\bf X} + \ov{\vb{\rho}}_{0},\; \ov{K}_{\rm gc}\}_{0} \}_{0} \equiv {\bf I}/m$ and 
\begin{equation}
\langle\chi_{2}\rangle \;\equiv\; \langle \ov{\bf A}_{1{\rm gc}}\bdot\{ \ov{\bf X} + \ov{\vb{\rho}}_{0},\; S_{1}\}_{0}\rangle \;=\; \left\langle \ov{\bf A}_{1{\rm gc}}\bdot\ov{\vb{\rho}}_{1{\rm gy}}\right\rangle.
\end{equation}

\end{document}